\documentclass[fleqn,usenatbib]{mnras}

\usepackage{newtxtext,newtxmath}
\usepackage[T1]{fontenc}
\usepackage{ae,aecompl}

\usepackage{graphicx}
\usepackage{amsmath,bm,mathtools}
\usepackage{booktabs}
\usepackage{array}
\usepackage{siunitx}
\usepackage{xcolor}

\title[three mode coupling bifurcation]{Resonant three-mode coupling in asteroseismology I: Bifurcations, chaos, and DAV white-dwarf outbursts}

\author[Guo \& Ogilvie]{Zhao Guo$^{1,2}$ and Gordon I. Ogilvie$^{2}$\\
$^{1}$Department of Applied Mathematics, School of Mathematics, University of Leeds, Leeds LS2 9JT, UK\\
$^{2}$Department of Applied Mathematics and Theoretical Physics, University of Cambridge, Cambridge CB3 0WA, UK
}

\date{Accepted Nov 6 666. Received Nov 6 666; in original form Nov 6 666}
\pubyear{2026}

\begin{document}
\label{firstpage}
\pagerange{\pageref{firstpage}--\pageref{lastpage}}
\maketitle

\begin{abstract}
Resonant three-mode coupling is a fundamental nonlinear process in stellar oscillations and other wave systems. We investigate the global dynamics of a linearly overstable parent mode coupled to two damped daughter modes. The principal control parameters are frequency detuning $\Delta$ and daughter damping rate $\Gamma$, normalized by the parent growth rate. We combine high-resolution maps of the largest Lyapunov exponent in the $(\Gamma,\Delta)$ plane with linear stability analysis, Floquet calculations, and numerical continuation. The resulting parameter space includes stable equilibria, simple and period-doubled limit cycles, chaotic attractors, formally unbounded solutions, and higher-period `shrimp' islands within chaotic regions. Successive period doublings lead to chaos, while boundaries of isolated periodic windows exhibit type-I intermittency.

At large $\Gamma$, the main bifurcation boundaries approach straight lines, indicating that the dynamics is governed primarily by $r=\Delta/(2\Gamma)$. We connect finite-$\Gamma$ bifurcations to an earlier one-dimensional return map. As $\epsilon\rightarrow0$, this map recovers the principal period-doubling sequence, the onset of chaos near $r\simeq0.38$, and the asymptotic positions of prominent periodic windows. We also examine the effects of unequal daughter damping rates.

Applying these results to sporadic outbursts in cool DAV white dwarfs, we find that estimated $g$-mode growth rates, damping rates, and detunings place resonant triplets in both periodic and chaotic regimes. Thus, instability of the three-mode equilibrium need not produce a regular limit cycle: deterministic chaos and intermittency may contribute to irregular outburst amplitudes and recurrence times.

\end{abstract}






\section{Introduction}


Resonant mode coupling is a generic consequence of weak nonlinearity in
oscillatory systems. When three waves or normal modes satisfy an approximate
frequency resonance,
\begin{equation}
    \omega_1 \simeq \omega_2+\omega_3 ,
\end{equation}
quadratic nonlinearities can produce a secular exchange of energy between
the modes on a timescale much longer than the oscillation period. Such
three-wave interactions occur in a wide variety of physical systems,
including stellar pulsations, internal and planetary waves in geophysical
fluids, plasma waves, and nonlinear optics
\citep[e.g.][]{SagdeevGaleev1969,Craik1988,Olbers2012,Kartashova2010}.

The corresponding amplitude equations follow naturally from a weakly
nonlinear multiple-scale expansion. In this approach, the solution is represented as a superposition of linear eigenmodes whose
complex amplitudes and phases evolve only on timescales much longer than
the oscillation periods. Projecting the nonlinear terms onto the linear
eigenmodes and averaging over the rapid oscillations removes
non-resonant contributions, leaving a reduced set of ordinary
differential equations (Amplitude Equations) that governs the slow exchange of energy and phase
among resonantly interacting modes.
This procedure has been developed extensively for stellar oscillations
\citep{Buchler1984,Goupil1994}, internal gravity waves
\citep{Bourget2013,Dauxois2018,Gururaj2020}, and Rossby and inertia--gravity waves
\citep{Raphaldini2019,Gururaj2025}. Amplitude equations can also be derived from averaged-Lagrangian or Hamiltonian formulations
\citep{Whitham1965,Dewar1972,Kumar1989,Kumar1994,VanHoolst1994b, Schenk2002}.
The same framework extends naturally from a single isolated triad to
networks of resonances. For example, two triads sharing one mode form a
five-wave system, or `butterfly' resonance cluster
\citep{Lvov2009,Gururaj2025}. Larger networks provide the finite-dimensional
building blocks of weak wave-turbulence theories \citep{Zakharov1992,Kartashova2010}, which is convenient to study energy transfer in fluids in the Fourier space \citep{Verma2019}.

Three-mode coupling has a long history in stellar pulsation theory.
Amplitude-equation formalisms were developed for both adiabatic and
non-adiabatic stellar oscillations
\citep{Dziem1982,Buchler1984,Goupil1994}, and resonant coupling has
subsequently been invoked in a variety of astrophysical contexts.
 It was shown that parametric excitation of damped daughter modes can strongly limit the observable amplitudes of
overstable $g$ modes of the Sun \citep{Dziembowski1983} and Slowly Pulsating B-stars \citep{Lee2012,VanB2024}, and $p$ modes of $\delta$ Scuti stars \citep{Dziem1982,Dziem1985,Dziem1988,Mourabit2025} and red-giant mixed modes \citep{Weinberg2021}. Similar parametric saturation has been studied for pulsating white dwarfs \citep{WuGoldreich2001}, rotating neutron stars and the $r$-mode instability \citep{Schenk2002,Arras2003}, and nonlinear
stellar tides \citep{Weinberg2012}. In each case an unstable or externally
driven parent mode can transfer energy to a pair of damped daughter modes
once the parent exceeds the parametric-instability threshold.

The three-mode amplitude equations possess rich
dynamics. This was
recognized early in plasma-physics studies of instability saturation.
\citet{Wersinger1980} showed that the resonant three-wave system can undergo
successive bifurcations from stationary solutions to periodic and irregular
motion. \citet{Bussac1982ThreeWave}, \citet{Bussac1982} and \citet{Hughes1990}
developed asymptotic descriptions of the strongly damped regime and showed
that the dynamics can be organized by a low-dimensional return map.
\citet{Meunier1982} and \citet{Bussac1982} further demonstrated
intermittent transitions between periodic and chaotic behaviour, including
Pomeau--Manneville \citep{PM1980} Type-I intermittency. These studies established that
even the minimal three-wave system can contain limit cycles,
period-doubling sequences, strange attractors, intermittent dynamics, and
embedded periodic windows. What remains less developed
in the stellar context is a global classification of these behaviours
in terms of the physically relevant detuning and damping parameters,
and a quantitative connection between this dynamical structure and
specific stellar mode triplets.

Related bifurcation phenomena also occur in much more complete models of
stellar pulsation. Nonlinear hydrodynamic calculations of Population-II
Cepheids revealed period-doubling cascades and transitions to deterministic
chaos as stellar parameters, particularly the effective temperature, are
varied \citep{BuchlerKovacs1987,KovacsBuchler1988}. \citet{Smolec2014} found a particularly rich bifurcation structure in
one-dimensional hydrodynamic models of BL Herculis stars (a subclass of Type-II cepheids), including
period-doubling cascades, periodic windows embedded in chaotic bands, intermittency, and interior crises. Thus many of the
dynamical phenomena familiar from low-dimensional nonlinear systems can
also arise in fully nonlinear stellar-pulsation calculations.

A particularly interesting astrophysical application is provided by
the outbursts observed in cool pulsating hydrogen-atmosphere white
dwarfs (DAVs, or ZZ Ceti stars) \citep{Bell2015,Hermes2015,Bell2016}. \citet{Luan18}, building on the
parametric-instability picture of \citet{WuGoldreich2001}, proposed
that these events arise from resonant coupling between an overstable
parent $g$ mode and a pair of damped daughter modes. While this
mechanism has usually been discussed in terms of unstable
equilibria and limit cycles, the corresponding three-mode system can
support a much wider range of nonlinear dynamics. DAV outbursts
therefore provide a natural astrophysical setting in which to examine
how periodic, period-doubled, intermittent, and chaotic behaviour is
organized across the detuning--damping parameter space.

In this paper we build on the earlier nonlinear-dynamics results and revisit the resonant three-mode amplitude
equations (Section~\ref{sec:amplitude}). We construct
two-dimensional Lyapunov maps of the frequency detuning and daughter-mode
damping rates, determine the principal equilibrium, Hopf and
period-doubling boundaries using linear stability and Floquet analyses, and
identify periodic windows, chaotic attractors (Section~\ref{sec:coupling}), and intermittency (Section~\ref{sec:intermittency}). We show that, at large daughter-to-parent damping ratio, the dynamics
is governed primarily by the ratio of detuning to daughter damping and
approaches the one-dimensional return-map description derived in
earlier work. This asymptotic map
reproduces the principal period-doubling sequence and the transition
to chaos seen in the full three-mode system (Section~\ref{sec:asymptotics}). Finally, we map
representative DAV mode parameters onto this bifurcation structure (Section~\ref{sec:application}) and
show that the three-mode outburst scenario encompasses not only regular
limit cycles but also deterministic chaotic and intermittent cycles. A discussion (Section~\ref{sec:discussion}) and conclusions (Section~\ref{sec:conclusions}) follow.

\section{Amplitude equations of 3-mode systems}

\label{sec:amplitude}

We briefly outline how the three-mode/wave amplitude equations arise from the nonlinear fluid equations. Resonant three-wave interactions occur in a wide range of physical systems, and the underlying mathematical structure of the amplitude equations is similar in different applications. In the following, we consider standing waves (normal
modes) in a non-rotating, inviscid, adiabatic,
self-gravitating star in hydrostatic equilibrium.

Introducing the Lagrangian displacement
$\boldsymbol{\xi}(\boldsymbol{r},t)$ and expanding the equations about the
equilibrium configuration gives, schematically,
\begin{equation}
    \rho_0\ddot{\boldsymbol{\xi}}
    +\mathcal{L}[\boldsymbol{\xi}]
    =
    \mathcal{N}_2[\boldsymbol{\xi},\boldsymbol{\xi}]
    +\mathcal{N}_3[\boldsymbol{\xi},\boldsymbol{\xi},
                   \boldsymbol{\xi}]
    +\cdots ,
    \label{eq:nonlinear_displacement}
\end{equation}
where $\mathcal{L}$ is the linear adiabatic oscillation operator, while
$\mathcal{N}_2$ and $\mathcal{N}_3$ collect the quadratic and cubic
nonlinearities, respectively \citep{Schenk2002}.   These nonlinear terms arise from the
advective acceleration and from the nonlinear perturbations of the
pressure, density, and gravitational potential.

The nonlinear displacement can then be expanded in the complete set of
linear eigenmodes,
\begin{equation}
    \boldsymbol{\xi}(\boldsymbol{r},t)
    =
    \sum_a q_a(t)\boldsymbol{\xi}_a(\boldsymbol{r}),
    \label{eq:mode_expansion_nonlinear}
\end{equation}
where the coefficients $q_a(t)$ describe the time-dependent modal
amplitudes and the linear modes satisfy   $\mathcal{L}[\boldsymbol{\xi}_a]
=\rho_0\omega_a^2\boldsymbol{\xi}_a $.

For the subsequent amplitude-equation formulation, we adopt an
energy normalization of the linear eigenmodes \citep{Weinberg2012}.  Specifically, the
eigenfunctions are scaled such that a complex modal amplitude of unit
magnitude corresponds to a fixed reference mode energy $E_0$,
\begin{equation}
    E_a = E_0 |q_a|^2 ,
\end{equation}
with $E_0=GM^2/R=1$ in the normalization adopted below.  This normalization
allows the conservative quadratic interaction of a resonant triad to
be described by a single dimensionless three-mode coupling coefficient
$\kappa_{abc}$.

Introducing slowly varying envelopes $A_j$ through
\begin{equation}
    q_j(t)=A_j(t)\,\mathrm{e}^{-\mathrm{i}\omega_j t} ,
\end{equation} 
and for a triad satisfying the near-resonance condition
\begin{equation}
    \Delta\omega
    \equiv
    \omega_a-\omega_b-\omega_c
    \simeq 0 ,
\end{equation}
we obtain \citep{Dziem1982, Schenk2002} 
\begin{align}
    \dot A_a+\gamma_aA_a
    &=
    2\mathrm{i}\omega_a\kappa_{abc}
    A_bA_c \,\mathrm{e}^{\mathrm{i}\Delta\omega t},
    \label{eq:slow_ae_a}\\
    \dot A_b+\gamma_bA_b
    &=
    2i\omega_b\kappa_{abc}^*
    A_aA_c^* \,\mathrm{e}^{-\mathrm{i}\Delta\omega t},
    \label{eq:slow_ae_b}\\
    \dot A_c+\gamma_cA_c
    &=
    2\mathrm{i}\omega_c\kappa_{abc}^*
    A_aA_b^* \,\mathrm{e}^{-\mathrm{i}\Delta\omega t}.
    \label{eq:slow_ae_c}
\end{align}
We rewrite the mode labels $a,b,c$ as $1,2,3$ respectively below. These equations are the energy-normalized form of the resonant
three-mode system obtained by projecting the quadratic nonlinear fluid equations onto the linear eigenmodes. Here $\gamma_j>0$ denotes linear damping, while a linearly
overstable mode has $\gamma_j<0$. These $\gamma_j$ are
\emph{amplitude} growth or damping rates. Since the mode energy
scales as $E_j\propto |q_j|^2$, the corresponding energy growth or damping rate $\gamma^E_j$ has magnitude $2|\gamma_j|$ \citep{WuGoldreich2001,Luan18}.

We label the overstable parent mode as mode 1 and the two damped
daughter modes as modes 2 and 3. Since $\gamma_1<0$, we define
the positive parent amplitude growth rate
$ \gamma_{\rm p}\equiv-\gamma_1>0$,
while $\gamma_2>0$ and $\gamma_3>0$ are the daughter amplitude
damping rates. We nondimensionalize the amplitude equations by
introducing
\[
    \tau=\gamma_{\rm p}t,\qquad
    \Gamma_2=\frac{\gamma_2}{\gamma_{\rm p}},\qquad
    \Gamma_3=\frac{\gamma_3}{\gamma_{\rm p}},\qquad
    \Delta=\frac{\Delta\omega}{\gamma_{\rm p}},
\]
together with the rescaled mode amplitudes 
$a_i = \frac{2\kappa_{abc}\sqrt{|\omega_j\omega_k|}}{\gamma_{\rm p}} |A_i|$. The resulting dimensionless amplitude equations are

\begin{align}
\frac{\mathrm{d}a_1}{\mathrm{d}\tau}&=a_1 + a_2 a_3 \sin{\phi}\\
\frac{\mathrm{d}a_2}{\mathrm{d}\tau}&=-\Gamma_2 a_2 -  a_3 a_1 \sin{\phi}\\
\frac{\mathrm{d}a_3}{\mathrm{d}\tau}&=-\Gamma_3 a_3 - a_1 a_2 \sin{\phi}\\
\frac{\mathrm{d}\phi}{\mathrm{d}\tau}&=\Delta -\left[ - \frac{a_2a_3}{a_1} + \frac{a_3a_1}{a_2}+\frac{a_1a_2}{a_3}\right ]\cos{\phi},
\label{AE}
\end{align} 
where $\phi$ is the three-mode coupling phase. Writing
the complex mode amplitudes as
$A_j=|A_j|\,\mathrm{e}^{-\mathrm{i}\theta_j}$, and choosing the eigenmode
phase convention such that the coupling coefficient
$\kappa_{abc}$ is real, we have
$ \phi=\theta_1-\theta_2 -\theta_3+\Delta\omega t$.

The system is of the form $\dot{\boldsymbol{x}}=\boldsymbol{F}(\boldsymbol{x})$ and its behaviour is controlled by parameters $(\Delta,\Gamma_2, \Gamma_3)$, or equivalently $(\Delta,\Gamma_2,s=\Gamma_3/\Gamma_2)$. We can interpret $\Delta$ as the three-mode detuning rate in units of the parent growth rate, $\Gamma_2$ and $\Gamma_3$ as the daughter damping rates in units of the parent growth rate, and $s$ as the ratio of daughter damping rates. Note that the three-mode coupling coefficient $\kappa_{abc}$ does not appear in the dimensionless system and so does not determine the behaviour of the system, although it does
affect the overall mode amplitudes.

\section{Three-mode coupling system behaviour and bifurcation}

\label{sec:coupling}

\begin{figure*}
\centering
\includegraphics[width=0.88\textwidth]{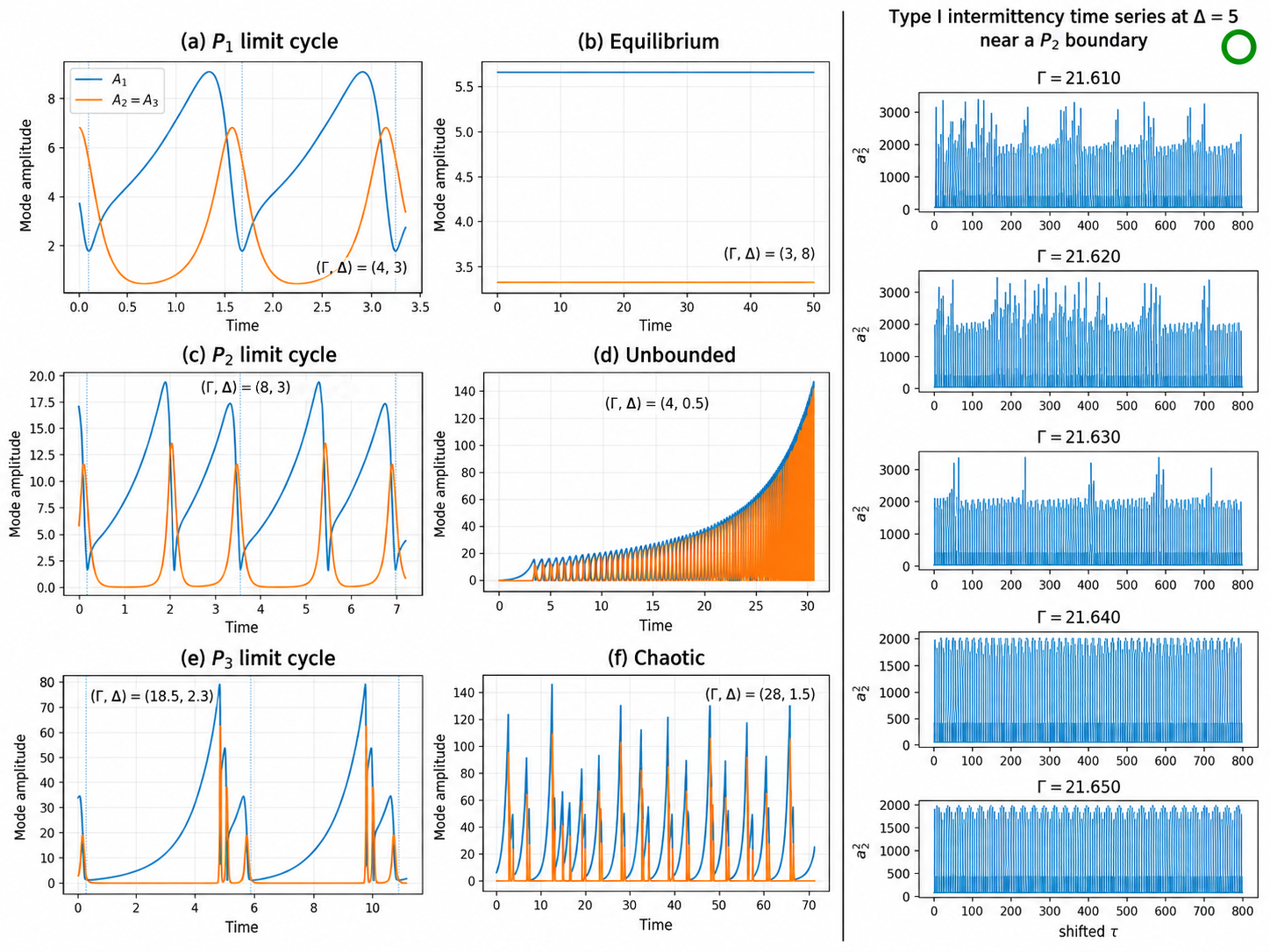}
\caption{Representative dynamics of the symmetric three-mode system
($\Gamma_2=\Gamma_3=\Gamma$).
The six panels on the left show the parent-mode amplitude $a_1$
(blue) and the identical daughter-mode amplitudes
$a_2=a_3$ (orange) for selected points in the $(\Gamma,\Delta)$
plane: (a) a $P_1$ limit cycle at $(4,3)$;
(b) a stable equilibrium at $(3,8)$;
(c) a period-doubled $P_2$ limit cycle at $(8,3)$; (d) a formally unbounded solution at $(4,0.5)$;
(e) a $P_3$ limit cycle at $(18.5,2.3)$; and
(f) a chaotic solution at $(28,1.5)$.
These points are marked by the corresponding labels in Fig.~\ref{fig:Del-Gam}.
The right column illustrates Pomeau--Manneville type-I
intermittency at fixed $\Delta=5$ near the boundary of the $P_2$
periodic region. As $\Gamma$ approaches the critical value
$\Gamma_{\rm c}\simeq21.6388$ from the chaotic side, increasingly
long nearly periodic (laminar) intervals are interrupted by
irregular bursts; beyond the boundary the trajectory settles onto
the stable $P_2$ limit cycle. Time $\tau$ is measured in units of the
inverse parent-mode growth rate. }
\label{fig:timeseries}
\end{figure*}

\begin{figure*}
\centering
\includegraphics[width=0.88\textwidth]{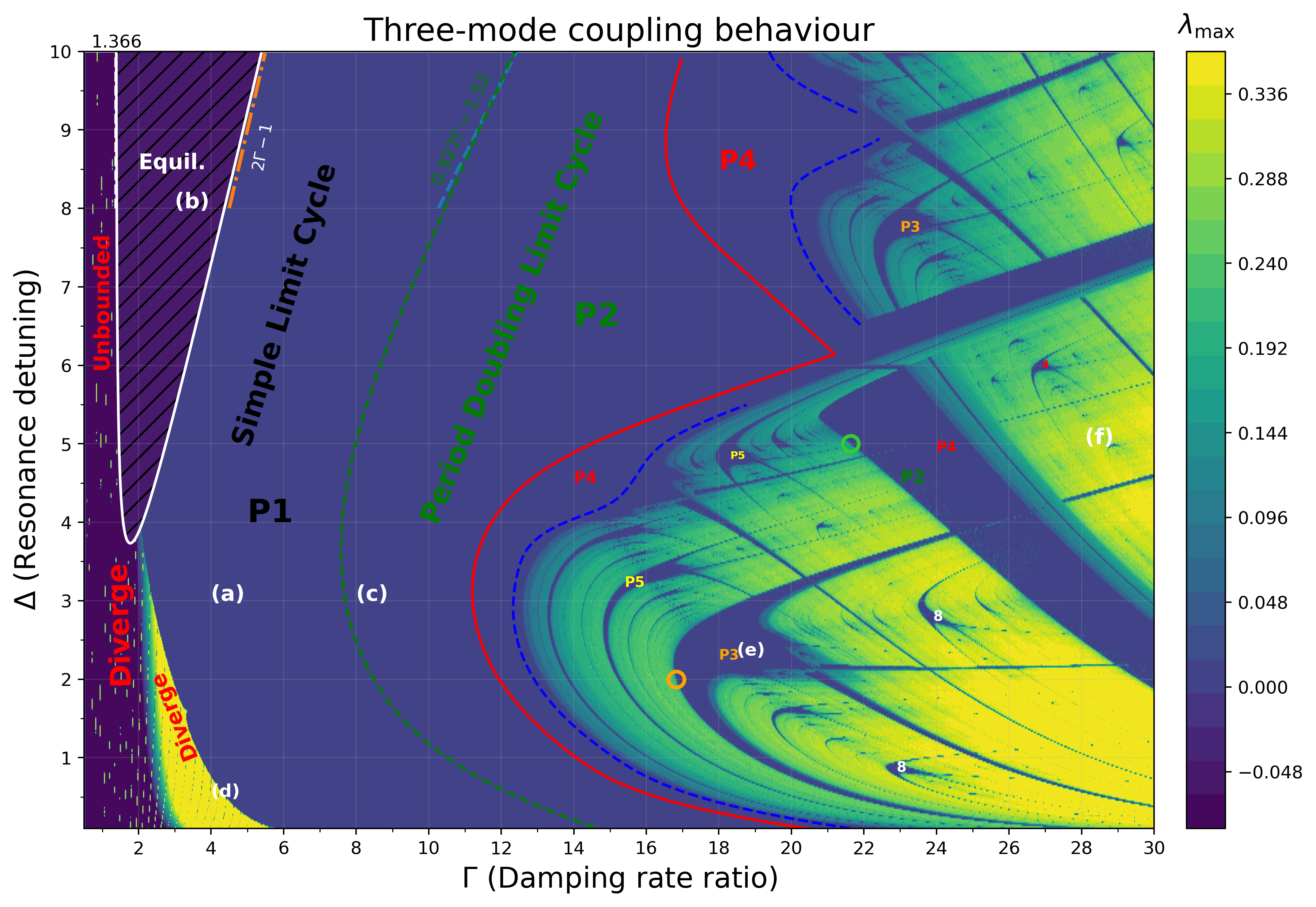}
\caption{Solution behaviour of the three-mode coupling system, represented by the
largest Lyapunov exponent $\lambda_{\rm max}$ in the $(\Gamma,\Delta)$
plane. The main dynamical regimes are an unbounded region at small
$\Gamma$ (labelled `Diverge' or `Unbounded'), a stable equilibrium
region enclosed by the white curve, periodic limit cycles denoted by
$P_n$, and a chaotic region (light green/yellow) containing numerous
shrimp-shaped periodic islands. The green dashed and red solid curves mark
the period-doubling transitions $P_1\rightarrow P_2$ and
$P_2\rightarrow P_4$, respectively. Labels (a)--(f) indicate the
parameter values of the six representative time series shown in
Fig.~\ref{fig:timeseries}. The green and orange circles mark the neighbourhoods in which the
intermittent behaviour is examined in Section 5 and Appendix~\ref{sec:boundary} (also in Fig.~\ref{fig:timeseries})}
\label{fig:Del-Gam}
\end{figure*}

\begin{figure*}
\centering
\includegraphics[width=0.88\textwidth]{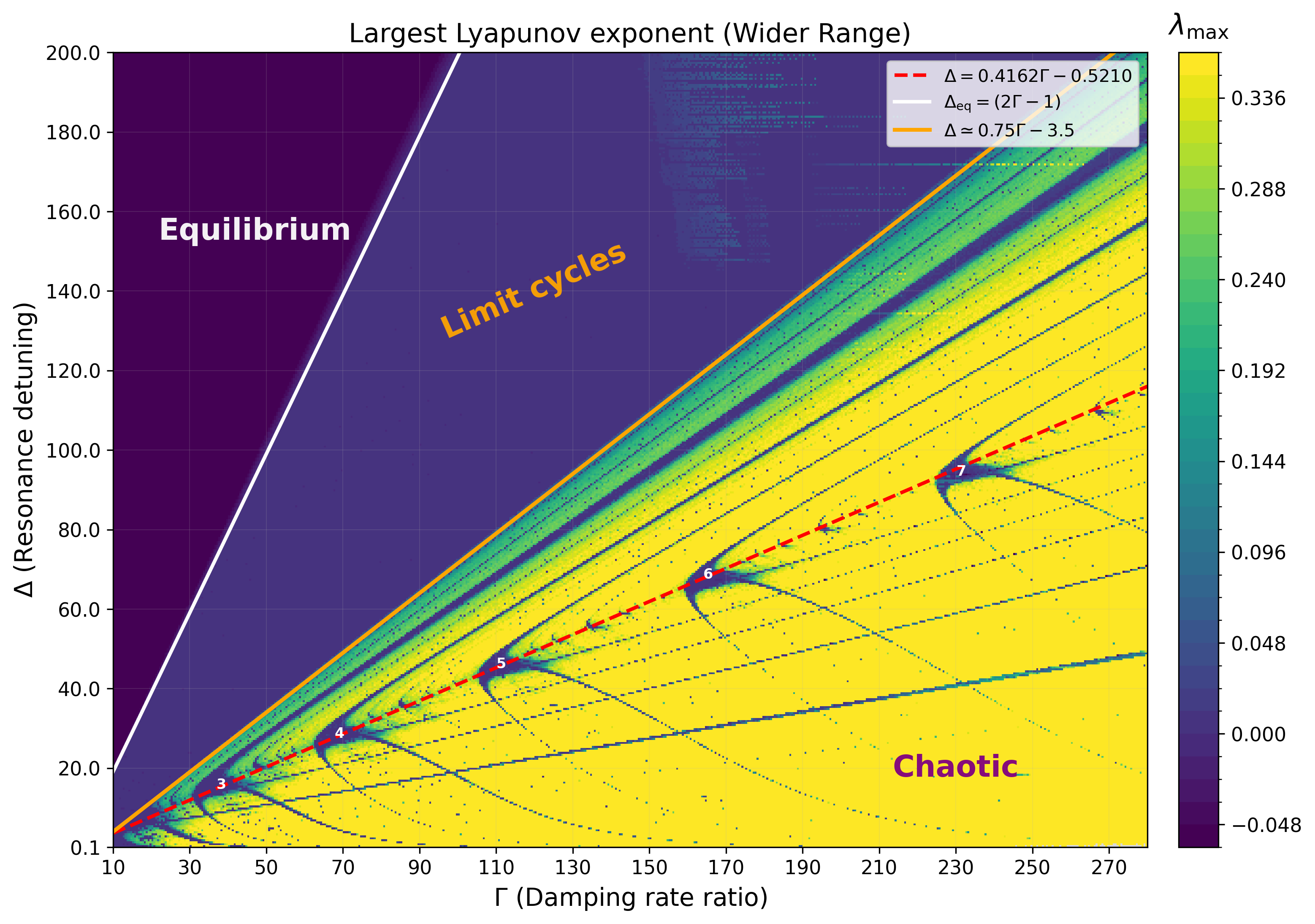}
\caption{Same as Fig.~\ref{fig:Del-Gam}, but over a wider region of the $(\Gamma,\Delta)$
parameter space. The white curve marks the transition from the stable
equilibrium region (dark purple) to periodic limit-cycle solutions, while
the orange curve marks the approximate transition from regular limit
cycles to chaotic dynamics (light green/yellow). Within the chaotic region,
a sequence of isolated shrimp-shaped periodic windows is visible; the
centres of the most prominent islands approximately follow the linear
relation indicated by the red dashed line. At large $\Gamma$, the principal
bifurcation boundaries become nearly linear in the $(\Gamma,\Delta)$ plane,
indicating that the dynamics depends primarily on the ratio
$\Delta/\Gamma$. This large-$\Gamma$ scaling is discussed in detail in
Section~4.}
\label{fig:shrimp}
\end{figure*}

The nonlinear behaviour of the three-mode system is illustrated first through representative time series in Fig.~1. We consider the identical-daughter case ($s=1$, $\Gamma_2=\Gamma_3=\Gamma$). The six examples in the left two columns correspond to the parameter combinations marked (a)--(f) in the enlarged Lyapunov map below (Fig. 2) and sample the main dynamical regimes encountered in the $(\Gamma,\Delta)$ plane. These include a stationary three-mode equilibrium (b), a simple ($P_1$) limit cycle (a), a period-doubled ($P_2$) limit cycle (c), a higher-period ($P_3$) limit cycle (e), an unbounded solution (d), and an irregular chaotic solution (f). The periodic solutions show the characteristic slow growth and rapid transfer of energy between the unstable parent (blue lines, labelled as $A_1$) and the damped daughters (orange lines, labelled as $A_2=A_3$), with the number and arrangement of bursts within one complete cycle changing across successive periodic windows. In a typical limit cycle, the parent mode grows slowly until it exceeds the threshold of parametric instability, after which the daughter modes grow rapidly into the nonlinear regime and quickly reduce the parent amplitude below threshold, after which the cycle repeats \citep{WuGoldreich2001,Pnigouras2015}. Note that in the chaotic regime (f), the burst amplitudes and recurrence intervals vary irregularly.

To characterize the long-term dynamics of the three-mode system across parameter space, we compute its Lyapunov exponents, which quantify the exponential growth or decay of infinitesimal perturbations to a trajectory. For a state vector $\boldsymbol{x}(t)$ satisfying Eq. \eqref{AE} with $\boldsymbol{x}(t)=\left (a_1,a_2,a_3,\phi \right )^T$ , an infinitesimal perturbation evolves according to the variational equation
\begin{equation}
    \dot{\delta\boldsymbol{x}}
    =
    \mathbf{J}(\boldsymbol{x})\,\delta\boldsymbol{x},
    \qquad
    \mathbf{J}
    =
    \frac{\partial\boldsymbol{F}}{\partial\boldsymbol{x}},
    \label{perturb_dx}
\end{equation}
and the corresponding Lyapunov exponent can be defined as
\begin{equation}
    \lambda
    =
    \lim_{t\rightarrow\infty}
    \frac{1}{t}
    \ln
    \frac{\|\delta\boldsymbol{x}(t)\|}
         {\|\delta\boldsymbol{x}(0)\|}.
\end{equation}

The rotational-transformed amplitude equations (Appendix~\ref{sec:rotational})
and the perturbed version were
integrated simultaneously to compute the Lyapunov exponents. All integrations started from
$\boldsymbol{x}_0=(0.5,0.5,0.5,0.5)$ and were carried
out over the dimensionless interval $0\leq\tau\leq2000$.
We used an adaptive-step fourth-order Runge--Kutta scheme
with typical timesteps $\delta\tau=(1$--$5)\times10^{-4}$.
The Lyapunov exponents were calculated following
\citet{Wolf1985}, by measuring the average exponential
growth or decay of small perturbations along the numerically integrated trajectories. A transient interval of $\tau_{\rm trans}=200$ was
excluded to allow the trajectory to approach its asymptotic attractor. For a given set of parameters, the asymptotic Lyapunov exponents are independent of the initial
conditions provided that the trajectories converge to
the same ergodic attractor. However, if multiple
attractors coexist, different initial conditions may
yield different Lyapunov exponents. Thus, our calculations characterize the attracting solution reached from the
adopted initial condition.

Of particular interest is the largest Lyapunov exponent,
$\lambda_{\max}$, since a positive value indicates exponential sensitivity to initial conditions and hence chaotic dynamics. Stable equilibria have $\lambda_{\max}<0$, whereas bounded periodic solutions of an autonomous system have $\lambda_{\max}=0$ because of the neutral direction associated with time translation. Chaotic attractors and also unstable equilibria are characterized by $\lambda_{\max}>0$. A two-dimensional dynamical map $\lambda_{\rm max}(\Gamma,\Delta)$ provides an efficient way to distinguish equilibrium, periodic, and chaotic regions and to reveal finer structures such as period-doubling sequences and isolated periodic windows embedded within the chaotic domain. Unbounded solutions are identified separately from the Lyapunov classification by monitoring the growth of the mode amplitudes during the integrations.

Fig.~\ref{fig:Del-Gam} shows a high-resolution view of the largest-Lyapunov-exponent map $\lambda_{\max}$ at relatively small $\Gamma$ and $\Delta$, with resolution of $0.01$ for both parameters. Several analytically or numerically determined bifurcation curves are overplotted. 

The white curve marks the loss of stability of the stationary
three-mode equilibrium through a Hopf bifurcation. The boundary
is obtained by linearizing the amplitude equations about the
stationary solution and considering perturbations governed by
Eq.~\eqref{perturb_dx}. Assuming a time dependence $\delta{\boldsymbol{x}}\propto \mathrm{e}^{\alpha\tau}$ gives the
characteristic equation \citep{Dimant2000,Pnigouras2015}
\begin{equation}
\alpha^4 + A \alpha^3 + B \alpha^2 + C \alpha + D = 0.
\end{equation}
The time unit is the reciprocal of the parent mode damping rate = $1/\gamma_{\rm p}$. The stationary solution becomes unstable and the system undergoes a Hopf bifurcation when a complex-conjugate pair of eigenvalues of the linearized system crosses the imaginary axis.
Setting $\alpha =\mathrm{i} \omega$, we can derive the frequency and period of the $P_1$ limit cycle at the Hopf bifurcation as:
$\omega = \sqrt{C/A}$, $P=2\pi/\omega$.

For the stationary solution to be linearly stable, the
Routh--Hurwitz criteria require the relevant Hurwitz determinants
to be positive. In particular,
$ABC-C^2-A^2D > 0 $.
At the Hopf bifurcation, this determinant vanishes, giving the
marginal-stability condition
\begin{equation}
ABC-C^2-A^2D = 0,
\label{stability}
\end{equation}
where $C = 4\Gamma_2 \Gamma_3  (1 + 3f^2)$, $f = \Delta/(\Gamma_2 + \Gamma_3 -1)$,
$A=2(\Gamma_2 + \Gamma_3 - 1)$,
$B= (\Gamma_2 + \Gamma_3 -1)^2+f^2(\Gamma_2 +\Gamma_3 -2\Gamma_2\Gamma_3 +2\Gamma_2 +2\Gamma_3+1)$, and 
$D=4\Gamma_2\Gamma_3(\Gamma_2+\Gamma_3-1)(1+f^2)$.

For equal damping rates with $\Gamma_3=\Gamma_2=\Gamma$, i.e.\ $s=1$, 
this can be simplified to:
\begin{equation}
    \Delta_{\rm P_1}
    =
    (2\Gamma-1)
    \left(
    \frac{2\Gamma^2-2\Gamma+1}
         {2\Gamma^2-2\Gamma-1}
    \right)^{1/2}
    \simeq 2\Gamma-1 .
    \label{twoGamma}
\end{equation}
This marks the boundary line between the equilibrium solutions and simple limit cycles $P_1$ (white line in Figs~\ref{fig:Del-Gam} and~\ref{fig:shrimp}).
There is a value of $\Gamma \approx 1.366$ (marked in Fig.~\ref{fig:Del-Gam}) below which the solution is always unstable, regardless of $\Delta$.

In Fig.~\ref{fig:Del-Gam}, the green curve denotes the first period-doubling boundary $P_1\rightarrow P_2$. It asymptotically (for large $\Gamma$) approaches a linear relation:
\begin{equation}
\Delta_{\rm P2} \approx 0.92\Gamma-1.5 .
\end{equation}
This boundary is determined from Floquet analysis of the stability of periodic solutions (detailed in Appendix~\ref{sec:floquet}) and verified by numerical continuation with \textsc{MatCont} \citep{Dhooge2003} , in agreement with \citet{Moskalik1985}. The red curve marks the subsequent period-doubling boundary $P_2\rightarrow P_4$, and the next level $P_4\rightarrow P_8$ boundary is shown as blue dashed lines. 
With a sequence of period doubling, the system eventually shifts to the chaotic regions (light green/yellow region).

Beyond the principal $P_1$, $P_2, P_4 ...$ domains, numerous higher-period windows are present inside the yellow chaotic region, including $P_3$, $P_4$, and other periodic islands embedded. Most of these isolated periodic regions have a four-legged shrimp shape. And we have labelled the periods of the solutions at the center of these `shrimps'. The labelled points (a)--(f) correspond directly to the representative time series shown in Fig.~\ref{fig:timeseries}, thereby linking the local temporal behaviour to its location within the global parameter-space structure. Notebly, (e) is a $P_3$ limit cycle at the center of the large shrimp near $\Gamma=18.5, \Delta=2.3$ and (f) shows a chaotic solution. 

Fig.~\ref{fig:shrimp} extends the calculation to much larger dimensionless damping rate $\Gamma \sim 280$ and reveals a striking organization of the dynamics. The white equilibrium boundary continues approximately as
$  \Delta_{P_1} \simeq 2\Gamma -1 $, while the transition between the broad regular limit-cycle region and the strongly chaotic region follows an approximately linear relation (orange line) in the $(\Gamma,\Delta)$ plane:
\begin{equation}
  \Delta_{\rm chaos} \approx 0.75\Gamma-3.5 
  \label{chaosline}
\end{equation}

In addition to the main period-doubling cascade, the chaotic region contains
a remarkable hierarchy of isolated periodic windows with the characteristic
`four-legged shrimp' morphology. The most prominent of these structures form a
well-defined sequence, labelled $3,4,5,6,7,\ldots$ in Fig.~\ref{fig:shrimp},
whose centers lie close to the empirical relation
\begin{equation}
    \Delta_{\rm shrimp} \simeq 0.416\,\Gamma - 0.52 ,
    \label{shrimp}
\end{equation}
shown by the red dashed line. This is the same organizing relation identified
by \citet{Rech2010}.
Moving along this line, the dynamics alternates between chaotic intervals and
stable periodic regions, while the characteristic period of the dominant
periodic structures increases successively. 

The large shrimps represent only the most visible level of this organization.
Between successive major structures, the apparently chaotic background
contains numerous smaller periodic islands with closely related shrimp-like
shapes. Successive magnification of these regions reveals further families of
periodic structures, including secondary period-adding sequences embedded
between the primary ones. \citet{Rech2010}, for example, found additional organized
structures between the principal $3$--$4$ and $5$--$6$ windows, and showed
that analogous patterns recur on progressively smaller parameter scales.
Different levels of this
hierarchy need not obey the same period-adding rule; besides the principal
sequence in which the period increases by unity, subsidiary sequences with
increments of two and other arrangements can occur. Moreover, near the
boundary of an individual shrimp, the stable periodic orbit can itself undergo
successive period-doubling bifurcations before entering the surrounding
chaotic region. Thus the yellow region in Fig.~\ref{fig:shrimp} should not be
regarded as uniformly chaotic, but rather as a highly structured mixture of
chaotic attractors and nested periodic windows extending over a wide range of
scales.

We discuss the the boundary lines $\Delta_{P_1},\Delta_{P_2},\Delta_{\rm chaos}$ vs. $\Gamma$ for the unequal-daughter case ($s =\Gamma_3/\Gamma_2 \neq 1$) in Appendix~\ref{sec:unequal}. 

\section{Large-$\Gamma$ asymptotics and a one-dimensional return map}

\label{sec:asymptotics}

Although the detailed parameter space contains a hierarchy of periodic windows, period-doubling structures, and chaotic bands, its large-scale organization is approximately controlled by the ratio of detuning to daughter damping. This motivates introducing
\begin{equation}
    r \equiv \frac{\Delta}{2\Gamma},
    \qquad
    \epsilon \equiv \frac{1}{\Gamma},
\end{equation}
and considering the asymptotic limit $\epsilon\rightarrow0$. In physical terms, $r$ measures the resonance mismatch relative to the daughter damping rate, while $\epsilon$ measures the separation between the parent-growth and daughter-damping timescales. The near-linearity of the principal boundaries at large $\Gamma$ (Fig.~\ref{shrimp}) therefore suggests that the dynamics approaches a reduced large-$\Gamma$ description in which $r$, rather than $\Delta$ and $\Gamma$ independently, becomes the dominant control parameter.

\begin{figure*}
\centering
\includegraphics[width=0.88\textwidth]{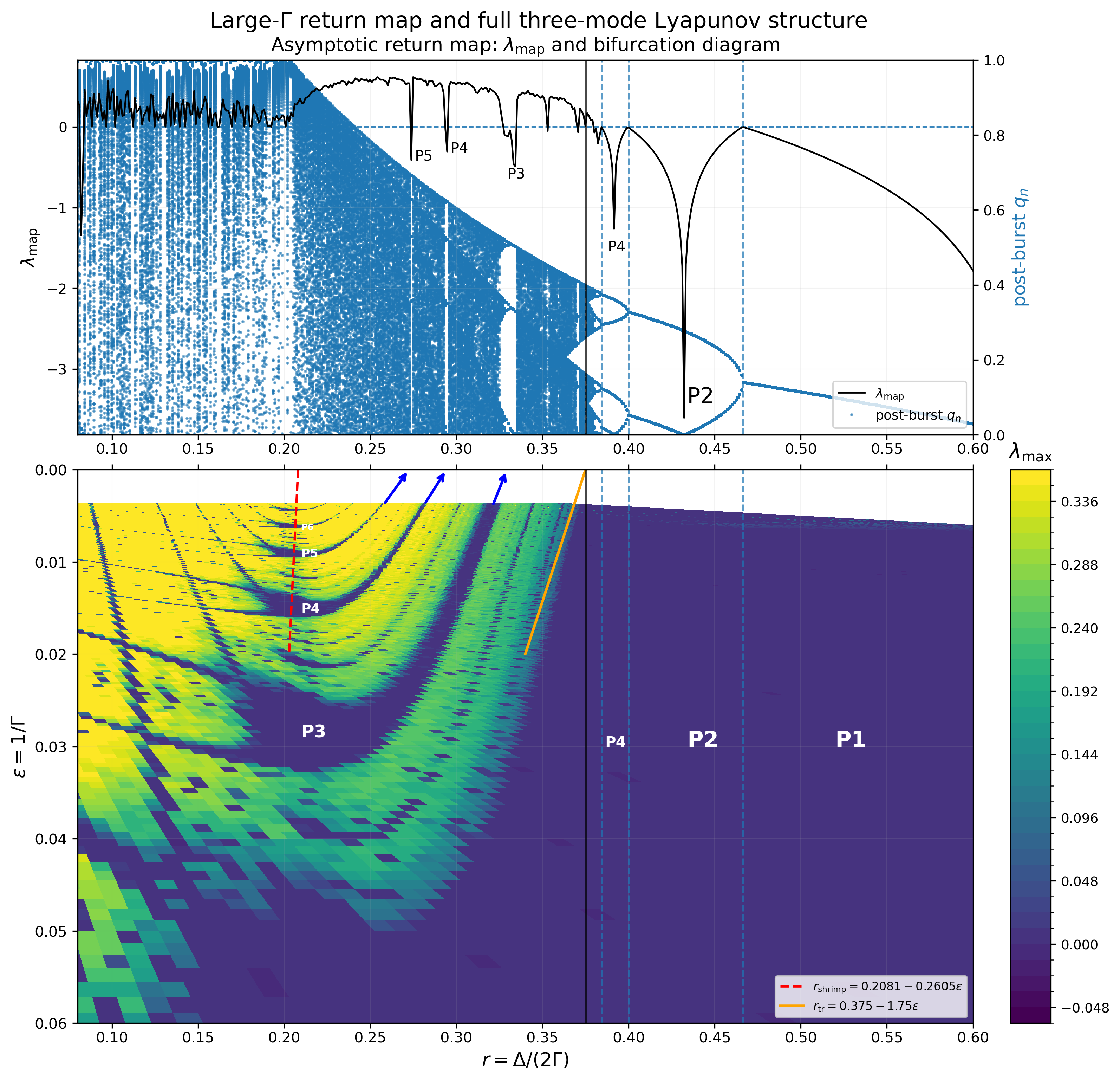}
\caption{\textbf{Lower:} Same Lyapunov exponents as shown in Fig.~\ref{fig:Del-Gam} but in the $(r=\Delta/(2\Gamma),\epsilon=1/\Gamma$) plane. The $\epsilon \rightarrow 0$ asymptotic behaviour is compared with the upper panel. \textbf{Upper:} Lyapunov exponent $\lambda_{\rm map}$ of the 1D return map (black) and the corresponding bifurcation diagram (blue). Dips in $\lambda_{\rm map}$ mark the major part of periodic regions. Vertical blue lines mark the transition points of periodic limit cycles $P_n \rightarrow P_{2n}$. This period-doubling sequence accumulates to the black vertical line at $r \sim 0.38$ which separates the periodic (dark blue) and chaotic (light green/yellow) region in the small-$\epsilon$ limit. The four-legged shrimp periodic islands in Fig.~\ref{fig:Del-Gam} now align nearly vertically (labelled as $P_3,P_4,P_5, ...$) and approach the red dashed line in the same limit. }
\label{fig:1dmap}
\end{figure*}

In the large-$\Gamma$ limit, the three-mode dynamics can be reduced to a one-dimensional return map \citep{Bussac1982ThreeWave,HughesProctor1992}. This limit is equivalent to the strongly damped three-wave problem
analysed by \citet{Hughes1990}. To make the correspondence
explicit, for equal daughter damping rates we introduce
\[
    x=\frac{X}{\Gamma},\qquad
    y=\frac{Y}{\Gamma},\qquad
    w=\frac{Z^2}{\Gamma^2},\qquad
    T=\Gamma\tau ,
\]
where $(X,Y,Z)$ are the Cartesian-like variables introduced in Eq. \eqref{phaseAE}.
The equations then become
\begin{align}
    \frac{\mathrm{d}x}{\mathrm{d}T}
      &= \epsilon x+2ry+w-2y^2,\\
    \frac{\mathrm{d}y}{\mathrm{d}T}
      &= \epsilon y-2rx+2xy,\\
    \frac{\mathrm{d}w}{\mathrm{d}T}
      &= -2w(1+x).
\end{align}
These are identical to Eq.~(1.4) of \citet{Hughes1990},
with their parameters related to ours by
$\gamma_{\rm HP}=\epsilon$ and $\delta_{\rm HP}=2r$.

\citet{Hughes1990} showed that, for
$\epsilon\ll1$, the dynamics consists of long slow phases
interrupted by rapid nonlinear bursts and can be reduced to a
one-dimensional return map. The return map describes how the outcome of a burst of parametric instability (in which the daughter modes grow rapidly to finite amplitude and reduce the parent amplitude well below threshold) sets up the cycle that follows.
Here we use their leading-order
singular map, rewritten in terms of the post-burst coordinate
$q=-x$:
\begin{equation}
    q_{n+1}=F_r(q_n)
    =B\!\left[S_r(q_n)\right],
    \qquad
    r=\frac{\Delta}{2\Gamma}.
\end{equation}
The slow map $S_r$ is determined implicitly by
\begin{equation}
    \Psi(q_+;r)=\Psi(q_-;r),
\end{equation}
where $q_-=q_n$, $q_+=S_r(q_n)>1$, and
\begin{equation}
    \Psi(q;r)
    =
    q-r\tan^{-1}\!\left(\frac{q}{r}\right)
    -\frac{1}{2}\ln(q^2+r^2).
\end{equation}

\begin{figure*}
\centering
\includegraphics[width=0.88\textwidth]{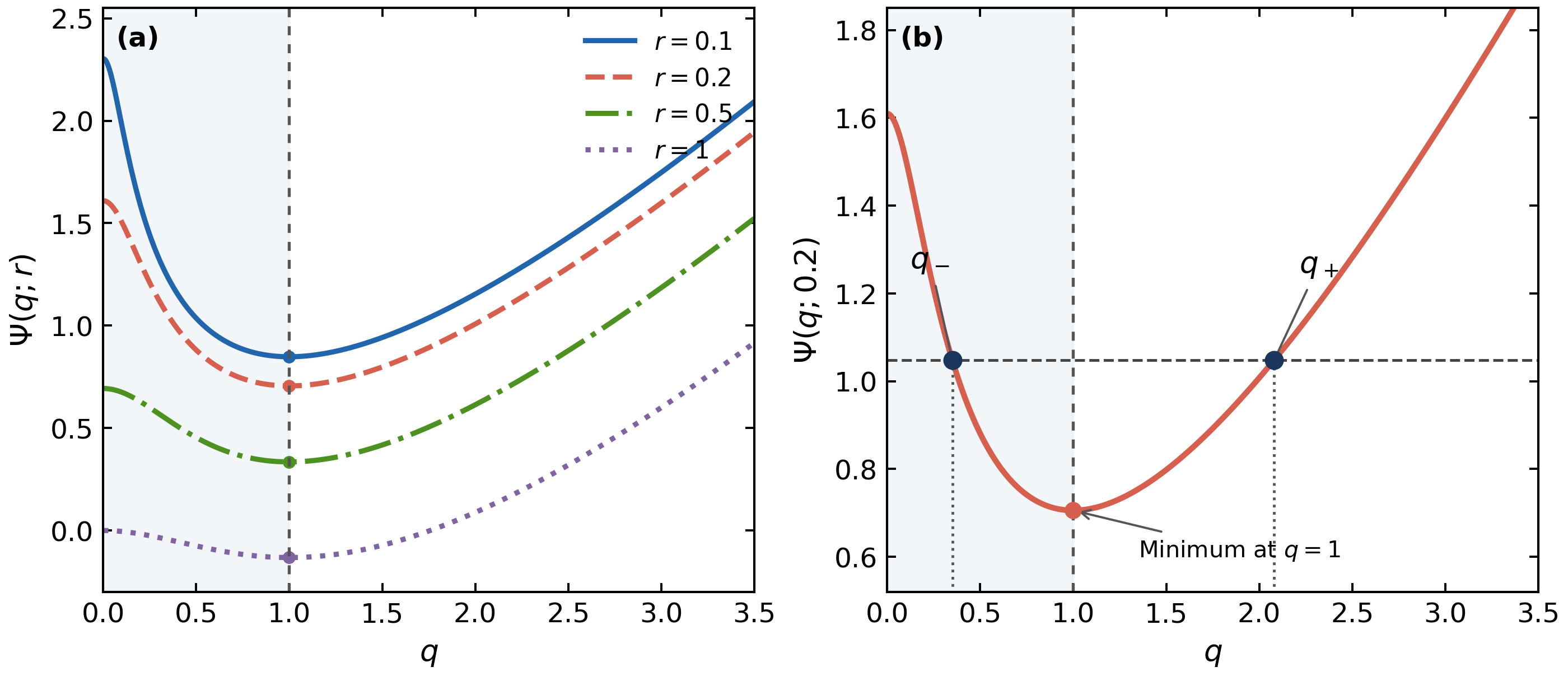}
\caption{\textbf{Left:} The slow map function $\Psi(q)$ for different values
of $r$. \textbf{Right:} Illustration of the implicit slow map $\Psi(q_+;r)=\Psi(q_-;r)$ for $r=0.2$. This example is for $q_-=0.35$ and $q_+\simeq2.0803$. }
\label{fig:psimap}
\end{figure*}

In Fig.~\ref{fig:psimap}, panel (a) shows the function $\Psi(q)$ for different values
of $r=\Delta/(2\Gamma)$. All curves have a minimum at $q=1$, separating the post-burst ($0<q<1$) and pre-burst ($q>1$) of the slow
evolution;
panel (b) illustrates the implicit slow map for $r=0.2$. For a given post-burst coordinate $q_-$, the corresponding
pre-burst coordinate $q_+$ is determined by
$\Psi(q_+;r)=\Psi(q_-;r)$.
Given $q_-=0.35$, the example shown has $q_+\simeq2.0803$.

During the subsequent rapid nonlinear burst, the trajectory is returned to the interval $0\leq q\leq1$. This reset can be represented
by a simple folding operation: values of $q$ are first reduced modulo
2, and values lying between 1 and 2 are then reflected about $q=1$.
Defining $\tilde q=q \bmod 2$, the burst map is therefore
\begin{equation}
B(q)=
\begin{cases}
\tilde q, & 0\leq \tilde q\leq1,\\
2-\tilde q, & 1<\tilde q<2.
\end{cases}
\end{equation}
Thus the slow phase determines how far the trajectory moves before the
burst, while the folding step gives the new post-burst value that starts
the next cycle. This 1D return map $q_{n+1}=F_r(q_n)$ is plotted in Appendix~\ref{sec:return}.

For each $r$, the bifurcation diagram is obtained by iterating $F_r$, discarding the first $1200$ iterates and retaining the following $300$. The map Lyapunov exponent follows from
\begin{equation}
    \lambda_{\rm map}
    =
    \lim_{N\rightarrow\infty}
    \frac{1}{N}
    \sum_{n=1}^{N}
    \ln\left|F_r'(q_n)\right|.
\end{equation}
Since $|B'|=1$ away from the folding cusps,
\begin{equation}
    \left|F_r'(q_n)\right|
    =
    \left|
    \frac{\Psi'(q_n;r)}
         {\Psi'(q_n^+;r)}
    \right|,
    \qquad
    \Psi'(q;r)=\frac{q(q-1)}{q^2+r^2}.
\end{equation}
Thus $\lambda_{\rm map}<0$ corresponds to stable periodic motion, while
$\lambda_{\rm map}>0$ indicates chaos; stability boundaries occur when
$\lambda_{\rm map}\rightarrow0$.

Fig.~\ref{fig:1dmap} compares the singular return-map limit with the full three-mode system. The upper panel shows the return-map bifurcation diagram and $\lambda_{\rm map}$ as functions of $r$, while the lower panel shows the largest Lyapunov exponent of the full 3-mode system in the $(r,\epsilon)$ plane, with
$ \epsilon=\Gamma^{-1} $.
Strong negative minima of $\lambda_{\rm map}$ identify maximally attracting periodic states. Prominent minima occur near
\begin{equation}
    r\simeq0.434\;(P_2),\qquad
    r\simeq0.388\;(P_4),
\end{equation}
these two locations mark the midpoint of the $P_2$ and $P_4$ limit cycle regions and match the full 3-mode result shown in the lyapunov colormap in the lower panel.

Within the chaotic regime, additional dips are shown at
\begin{equation}
    r\simeq0.3296\;(P_3),\qquad
    0.2966\;(P_4),\qquad
    0.276\;(P_5).
\end{equation}
The periodic `shrimp leg' structures of the finite-$\epsilon$ system bend toward these values as $\epsilon\rightarrow0$ (indicated by the blue arrows in the lower panel), suggesting that the minima of the 1-D return map act as asymptotic anchors for the most stable parts of the finite-$\Gamma$ periodic windows. 

The return map also reproduces the principal period-doubling sequence. The approximate
$P_1\rightarrow P_2$, $P_2\rightarrow P_4$, and $P_4\rightarrow P_8$ boundaries occur at
\begin{equation}
    r_{P_2}\simeq 0.4662,\qquad
    r_{P_4} \simeq 0.400,\qquad
    r_{P_8} \simeq 0.3847,
\end{equation}
respectively. At these boundaries, the corresponding periodic orbit loses stability and $\lambda_{\rm map}$ passes through zero. And this sequence accumulates near $r _{\rm chaos, map}\simeq 0.38$. This agrees well with the large-$\Gamma$ limit of the empirical regular-to-chaotic transition of the full system in \eqref{chaosline} (black vertical line in both upper and lower panel of Fig.~\ref{fig:1dmap}),

\begin{equation}
    r_{\rm chaos}\rightarrow0.375.
\end{equation}
The complicated two-parameter structure of the full three-mode equations is therefore organized, in the large-$\Gamma$ limit, by the much simpler one-dimensional return map: period-doubling boundaries approach zeros of $\lambda_{\rm map}$, while periodic shrimp cores approach its deep negative minima.

The red dashed line in the lower panel of Fig.~\ref{fig:1dmap} shows the numerical fit to the centres of the shrimp structures Eq.~(21). As $\epsilon \rightarrow 0$, the fitted relation asymptotically approaches $r \simeq 0.21$. Interestingly, this limiting value closely coincides with the point at which the return-map Lyapunov exponent, $\lambda_{\rm map}$, changes curvature in the upper panel. The same value of $r$ also marks a noticeable change in the density and organisation of the branches in the corresponding bifurcation diagram, suggesting that the asymptotic location of the shrimp cores is tied to a structural transition in the large-$\Gamma$ return map.


\section{Type-I Intermittency near a periodic boundary point}

\label{sec:intermittency}

The right-hand panels of Fig.~\ref{fig:timeseries} examine the dynamics close to selected periodic-window boundaries in greater detail. In particular, the time series near the $P_2$ limit cycle boundary (at $(\Gamma,\Delta)=(21.64,5)$ as we vary $\Gamma$ at fixed $\Delta=5$) show long intervals of nearly periodic behaviour interrupted by irregular excursions, characteristic of intermittency. We refer to these nearly periodic intervals between successive irregular bursts as \emph{laminar phases}, and their lengths as the laminar durations. The corresponding regions in parameter space are indicated by the green circle in Fig. 2. The orange circle marks the neighbourhood of a shrimp-shaped $P_3$ periodic region near $(\Gamma,\Delta)=(16.83, 2)$.

Both neighbourhoods exhibit Pomeau--Manneville Type-I intermittency \citep{PM1980,Man1990}, associated with a tangent (saddle-node) bifurcation of a periodic
orbit. A characteristic signature is that the mean laminar duration increases as the critical parameter $\Gamma_c$ is approached, following the scaling $\langle L\rangle\propto|\Gamma-\Gamma_c|^{-1/2}$. We provide a detailed numerical verification in the Appendix~\ref{sec:boundary}.

\section{Application to pulsating white dwarf: outbursts in ZZ Ceti}
\label{sec:application}


Outbursts have recently emerged as a distinctive nonlinear phenomenon in cool pulsating hydrogen-atmosphere white dwarfs (DAVs, or ZZ Ceti stars). They were first identified in the \textit{Kepler} light curve of KIC~4552982, where brightenings lasting $4$--$25\,{\rm h}$ recur on a characteristic timescale of a few days \citep{Bell2015}. Similar behaviour was subsequently found in PG~1149+057 \citep{Hermes2015} and several additional cool DAVs observed by \textit{K2} \citep{Bell2016,Bell2017,Hermes2026}. These objects lie close to the cool (red) edge of
the DAV instability strip and typically exhibit outbursts lasting several to
tens of hours, separated by days, with energies of order
$10^{33}$--$10^{34}\,{\rm erg}$. A promising explanation is nonlinear
parametric coupling between an overstable parent $g$ mode and a pair of
linearly damped daughter modes \citep{WuGoldreich2001}. Building on this
picture, \citet{Luan18} showed that sufficiently resonant
three-mode interactions can undergo limit cycles in which energy accumulated
slowly in the parent mode is rapidly transferred to and dissipated by the
daughter modes. Such cycles naturally reproduce the observed separation
between the hours-long outburst duration and days-long recurrence time, as
well as the characteristic outburst energies and their prevalence near the
red edge of the DAV instability strip.

The nonlinear dynamics identified in this work provide a more detailed
interpretation of the DAV outburst scenario proposed by \citet[hereafter LG18]{Luan18}.  In their picture, an overstable parent $g$ mode is resonantly coupled to two linearly damped daughter modes,
with
\begin{equation}
    \omega_{\rm d1}+\omega_{\rm d2}
    \simeq \omega_{\rm p},
\end{equation}
and, for nearly identical daughters,
$\omega_{\rm d1}\simeq\omega_{\rm d2}\simeq\omega_{\rm p}/2$.

To distinguish the energy growth/damping rates used by
\citet{WuGoldreich2001} and LG18 from the amplitude rates used
in our equations, we denote the former by a superscript $E$.
Thus
\[
    \gamma_{\rm p}^{E}=2\gamma_{\rm p},\qquad
    |\gamma_{\rm d}^{E}|=2\gamma_{\rm d},
\]
where $\gamma_{\rm p}>0$ and $\gamma_{\rm d}>0$ are the
corresponding amplitude growth and damping rates, and

\begin{equation}
    \Gamma =
    \frac{|\gamma_{\rm d}|}{|\gamma_{\rm p}|}=\frac{\gamma^E_{\rm d}}{\gamma^E_{\rm p}},
    \qquad
    \Delta =
    \frac{|\delta\omega|}{|\gamma_{\rm p}|}=\frac{2|\delta\omega|}{\gamma^E_{\rm p}}.
\end{equation}  
Hence
\begin{equation}
    \frac{|\delta\omega|}{|\gamma^E_{\rm d}|}
    =
    \frac{\Delta}{2\Gamma} = r,
\end{equation}
and the classical condition $|\delta\omega|<|\gamma^E_{\rm d}|$
corresponds to
\begin{equation}
    \Delta < 2\Gamma,{\rm or \ } r <1
\end{equation}
consistent with the large-$\Gamma$ limit of our exact equilibrium stability boundary $ \Delta_{\rm P_1} \simeq 2\Gamma-1$ in Eq. \eqref{twoGamma}.

LG18 associated outbursts with the equilibrium-unstable regime
$|\delta\omega|<|\gamma^E_{\rm d}|$, in which the parent grows slowly,
excites the daughters above the parametric-instability threshold,
and subsequently transfers its energy rapidly to them. The slow recovery is associated roughly with the parent growth time, while the rapid burst is controlled by daughter damping: $t_{\rm rec} \sim 1/\gamma^{E}_p$ and $t_{\rm burst} \sim 1/|\gamma^{E}_d|$. Radiative damping of the daughters converts this mechanical energy into heat, producing an outburst, after which the parent starts to grow
again. 

A useful estimate of the physically relevant range of $\Gamma$ can
be obtained from the linear growth and damping rates shown in
Fig.~4 of LG18.  Although the parent is linearly
overstable whereas the daughters are damped, the parent growth rate
can still be related to the underlying radiative-damping scale.
For a DAV $g$ mode, the ratio of convective driving to radiative
damping is approximately
\begin{equation}
    \frac{\gamma^E_{\rm cvz}}
         {|\gamma^E_{\rm rad}|}
    \simeq
    \frac{2(\omega\tau_c)^2}
         {1+(\omega\tau_c)^2},
\end{equation}
where $\tau_c$ is the effective thermal timescale of the convection zone.
Writing $x_{\rm p}=\omega_{\rm p}\tau_c$, the net growth rate of an
overstable parent is therefore
\begin{equation}
    \gamma^E_{\rm p}
    =
    |\gamma^E_{\rm rad,p}|
    \frac{x_{\rm p}^2-1}{x_{\rm p}^2+1}.
\end{equation}
The peak-driven modes in Fig.~4 of LG18 have
$x_{\rm p}\simeq6$--$9$, so that
\begin{equation}
    \gamma^E_{\rm p}
    \simeq (0.95\mbox{--}0.98)
    |\gamma^E_{\rm rad,p}|.
\end{equation}
Thus, for well-overstable parent modes away from marginal stability,
the magnitude of the net parent growth rate remains close to the
underlying radiative-damping scale.  This motivates the
order-of-magnitude estimate
\begin{equation}
    \Gamma
    =
    \frac{|\gamma^E_{\rm d}|}{\gamma^E_{\rm p}}
    \simeq
    C_{\rm p}
    \frac{|\gamma^E_{\rm rad,d}|}
         {|\gamma^E_{\rm rad,p}|},
    \qquad
    C_{\rm p}
    =
    \frac{x_{\rm p}^2+1}{x_{\rm p}^2-1}
    \simeq1,
\end{equation}
provided that the parent is not close to marginal overstability.

Over a restricted range of radial order and frequency, the
radiative damping may be represented phenomenologically as
\begin{equation}
    |\gamma^E_{\rm rad}|
    \propto \omega^{-q_{\rm eff}},
\end{equation}
so that for nearly symmetric daughter pairs,
$\omega_{\rm d}\simeq\omega_{\rm p}/2$,
\begin{equation}
    \Gamma
    \simeq
    \left(
    \frac{\omega_{\rm p}}{\omega_{\rm d}}
    \right)^{q_{\rm eff}} \simeq  2^{q_{\rm eff}} .
\end{equation}

The simplest WKB radiative-diffusion argument gives a shallower frequency dependence, approximately
$|\gamma^E_{\rm rad}|\propto\omega^{-2}$ for a fixed propagation
cavity, since $k_r\propto\omega^{-1}$.  In cool DAVs, however, the
outer turning point and the degree of non-adiabaticity vary strongly
with frequency, steepening the effective dependence.  The g-mode damping/growth rates in
Fig.~4 of LG18 suggest
$\Gamma\sim10$--$30$ for many same-$\ell$ parent--daughter
combinations, corresponding to
\begin{equation}
    q_{\rm eff}
    \simeq3.3\mbox{--}4.9.
\end{equation}
Thus\footnote{\citet{Burkart2012} also find that $g$ modes in A-type main sequence star KOI-54 roughly satisfy $\gamma_{\rm nl} \propto (l/\omega_{\rm nl} )^4$ for the same reason (changing turning point) as for DAV $g$ modes.} $q_{\rm eff}\simeq 4$, for
which $\Gamma\simeq 16$, provides a useful representative estimate for the possible $\Gamma$ values of DAV white dwarf $g$ modes responsible for the outbursts. The explicit examples of Luan \& Goldreich give damping-to-growth ratios of order
$\Gamma\simeq20$--$30$, reinforcing the conclusion that the
region $\Gamma\sim10$--$30$ is particularly relevant for cool DAVs.


The detuning provides the second important constraint.
The expected detuning of the dynamically important daughter pair can be
estimated from the density of high-order $g$ modes.  For fixed angular
degree, the radial order satisfies the asymptotic relation $n \propto P \propto \omega^{-1}$, which
implies a local frequency spacing
\begin{equation}
    \delta\omega_{\rm d}
    \sim
    \left|\frac{\mathrm{d}\omega}{\mathrm{d}n}\right|
    \sim
    \frac{\omega_{\rm d}}{n_{\rm d}} .
\end{equation}
For a near-resonant daughter pair with
$\omega_{\rm d1}\simeq\omega_{\rm d2}\simeq\omega_{\rm p}/2$,
changing one daughter by one radial order changes the pair-sum frequency
by approximately $\omega_{\rm p}/(2n_{\rm d})$.  A characteristic
nearest mismatch within a single such mode sequence is therefore
\begin{equation}
    |\delta\omega|
    \sim
    \frac{\omega_{\rm p}}{4n_{\rm d}},
    \label{dw}
\end{equation}
which corresponds, in our normalization, to
\begin{equation}
    \Delta
    =
    \frac{2|\delta\omega|}{\gamma^E_{\rm p}}
    \sim
    \frac{\omega_{\rm p}}
         {2n_{\rm d}\gamma^E_{\rm p}} .
\end{equation}
For the parameters of Fig.~6 of \citet{Luan18},
$P_{\rm p}\simeq12.1\,{\rm min}$,
$n_{\rm d}\simeq58$--$59$, and
$\gamma^E_{\rm p}\simeq5.1\times10^{-6}\,{\rm s^{-1}}$, this estimate
gives $\Delta\sim15$ for a single daughter sequence.

In practice, however, many daughter pairs satisfying the angular selection
rules can compete for a given parent, and the pair with the lowest
parametric-instability threshold is preferentially selected.  Searching
over several independent candidate pairs therefore reduces the minimum
frequency mismatch by a factor of several or more.  
The dynamically relevant pair is
selected from many daughter pairs satisfying the angular selection
rules and having large coupling coefficients. \citet{WuGoldreich2001}
estimated statistically that the minimum frequency mismatch among
such pairs scales approximately as
\begin{equation}
    |\delta\omega|_{\rm min}
    \sim
    \frac{\omega_{\rm p}}
         {\ell_{\rm d}^{\,3} n_{\rm p}^{\,2}} .
\end{equation}
For low-degree DAV $g$ modes, this estimate can be
substantially smaller than the single-sequence estimate in
Eq.~\eqref{dw}. For $\ell_{\rm d}=1$ and $n_{\rm p}\sim20$--$30$, the
reduction is typically a factor of a few\footnote{The example in Fig. 6 of LG18 with parameters $\ell_{\rm d}=2$, $n_{\rm p}=28$ give a reduction by a factor of $\sim27$, corresponding to $\Delta\sim0.5$, close to the actual value of $0.46$.}. LG18 further found that the daughter pair giving the lowest
parametric-instability threshold typically satisfies
$|\delta\omega|\lesssim |\gamma_{\rm d}^{E}|$, and that for
sufficiently high-order parent modes
$|\delta\omega|\ll|\gamma_{\rm d}^{E}|$ can occur even for
$\ell_{\rm d}=1$. In our variables this corresponds to $ r \lesssim 1$,
with strongly resonant high-order triplets having $r\ll1$.
For the DAV range $\Gamma\sim10$--$30$, such optimal daughter pairs
therefore naturally populate values of $\Delta$ from below unity to
of order ten. 

The examples in LG18 corresponding to their Figs.~6 and 7 have approximately
\begin{equation}
    (\Gamma,\Delta)_{\rm LG18}\simeq(20.9,0.46)
    \quad\mathrm{and}\quad
    (30.5,4.55),
\end{equation}
respectively, corresponding to
$r \simeq0.01$ and $0.075$.  These values lie deep inside the equilibrium-unstable part ($r \lesssim 1$) of our parameter plane.

Our Lyapunov map provides a finer dynamical classification of the equilibrium-unstable region,
it resolves this broad equilibrium-unstable
region into qualitatively different nonlinear states.  The transition to strongly positive Lyapunov exponents is approximately
described by the empirical relation
\begin{equation}
    \Delta_{\rm chaos}
    \simeq 0.75\Gamma-3.5,
\end{equation}
which, for $\Gamma\sim10$--$30$, corresponds approximately to
\begin{equation}
    \frac{|\delta\omega|}{|\gamma^E_{\rm d}|}
    =
    \frac{\Delta_{\rm ch}}{2\Gamma}
    = r_{\rm chaos}
    \sim 0.38 .
\end{equation}
Thus, the two examples in \citet{Luan18} are actually well inside the chaotic region since $r < r_{\rm chaos}$. This explains the why their trace plots (Fig. 6 and 7) show aperiodic behaviour instead of strictly periodic or quasiperiodic. Thus the classical condition $|\delta\omega|<|\gamma^E_{\rm d}|$ (or $r <1$) encompasses a much richer set of
dynamical behaviours: moderately resonant triplets may exhibit
regular nonlinear cycles, whereas more strongly resonant systems
can enter a chaotic regime in which the amplitudes and recurrence
times vary irregularly from cycle to cycle.  This provides a natural
extension of the `limit-cycle' description of LG18, and particularly relevant to the aperiodic recurrence of observed DAV outbursts. 

The chaotic region itself contains narrow periodic windows organized
into the characteristic four-legged `shrimp' structures.  The
principal sequence approximately follows
\begin{equation}
    \Delta_{\rm shrimp}
    \simeq0.4162\Gamma-0.5210,
\end{equation}
which at large $\Gamma$ corresponds to
\begin{equation}
r=   \frac{|\delta\omega|}{|\gamma^E_{\rm d}|}
    \simeq
    \frac{0.4162}{2}
    \simeq0.21.
\end{equation}
Consequently, DAV triplets with very similar damping ratios and
detunings can exhibit qualitatively different outburst patterns:
chaotic regions correspond to irregular sequences of energy-transfer
events, while neighboring shrimp windows recover stable periodic
cycles of different multiplicities.  The $(\Gamma,\Delta)$
Lyapunov map therefore refines the original DAV outburst picture by
showing that the condition
$|\delta\omega|<|\gamma^E_{\rm d}|$ does not define a single type of
limit cycle, but instead contains simple periodic cycles, period-doubled states, chaotic attractors, and embedded periodic windows.

In this interpretation, the observed diversity of DAV outburst recurrence patterns may arise naturally from modest differences in daughter damping rates and resonance detuning among otherwise similar three-mode systems. Observed DAV outbursts indeed show different levels of periodicity, from nearly regular to very irregular outburst durations and spacings \citep{Bell2015}, this agrees very well with the behaviour in the $(\Gamma,\Delta)$ parameter space.


\section{Discussion}
\label{sec:discussion}

The current work on three-mode coupling system does not consider the effect of noise. \citet{Hughes1990} showed that the strongly damped three-wave system can be highly sensitive to weak noise. Even very weak stochastic forcing can therefore alter the escape from this slow phase and substantially modify the burst amplitude and recurrence period.


A Lyapunov map constructed from one initial condition identifies the attractor reached
from that particular part of phase space, but does not by itself exclude multistability.  Coexisting periodic and chaotic attractors, if present, would introduce an additional dependence on the initial modal amplitudes and phases and on their basins of attraction.  




An isolated triad is only the
simplest component of a stellar nonlinear mode network.  A given overstable mode generally has many potential daughter pairs \citep{Nowakowski2005}, while a
daughter of one resonance may participate in additional resonances.
Such interconnected triads can form higher-dimensional mode networks.  This is
particularly relevant in stars with dense oscillation spectra.  Four-mode and multi-triad systems \citep{Venumadhav2014, Weinberg2016} may therefore generate
additional modulation, switching between resonances, or higher-dimensional chaos \citep{Miranda2012}.  

Rotation introduces further possibilities.  Rotational splitting creates closely spaced multiplets and therefore increases the number
of potential resonances. It also naturally generates near-resonance four mode coupling $2\omega_0 \approx \omega_1 + \omega_2$ \citep{Buchler1995}. While differential rotation or a mean flow can modify the mode frequencies and damping rates themselves, and wave-generated Reynolds stress and nonlinear sources can feedback to the mean flow \citep{Grimshaw1984, Buhler2014, Guo2023}.  Extending the amplitude equations to include additional resonant modes \citep{Nowakowski2001} and a slowly evolving mean flow with mean flow \citet{Voelker2021} would therefore provide a natural next step for applications to rotating and nonlinear pulsators. It is also interesting to extend the current work to include the tidal forcing to the parent mode \citep{Weinberg2012}.

Finally, our DAV application is primarily a classification of the
possible nonlinear dynamics. Quantitative predictions of observable
outbursts additionally require realistic mode eigenfunctions, coupling
coefficients, damping rates, and thermal transport linking daughter-mode
dissipation to the emergent flux \citep{WuGoldreich2001,Luan18}.
Real DAVs may also involve several simultaneously active resonances.
Combining the present bifurcation analysis with mode properties from
realistic white-dwarf models could therefore connect the dynamical
regime of a given triplet with observable outburst energies, durations,
and recurrence times. More broadly, a systematic survey of resonant
triads, their coupling strengths, and their nonlinear behaviour across
different classes of pulsating stars would be valuable.
\section{Conclusions}
\label{sec:conclusions}

We have investigated the nonlinear dynamics of resonant three-mode
coupling between one linearly overstable parent mode and two damped
daughter modes. The dynamics is controlled
primarily by the frequency detuning $\Delta$ and the daughter-to-parent
damping-rate ratios $\Gamma_2$ and $\Gamma_3$. For identical daughters,
$\Gamma_2=\Gamma_3=\Gamma$, we constructed two-dimensional maps of the
largest Lyapunov exponent in the $(\Gamma,\Delta)$ plane and combined
these with linear stability analysis, Floquet calculations, and numerical
continuation. The resulting parameter space contains stable equilibria,
simple and period-doubled limit cycles, chaotic attractors, unbounded
solutions, and numerous isolated higher-period windows embedded within
the chaotic region.

Several simple relations organize the 
bifurcation structure. The stationary three-mode equilibrium loses
stability through a Hopf bifurcation at
$\Delta_{\rm P_1}\simeq2\Gamma-1$ for large $\Gamma$, while the first
period-doubling boundary approaches
$\Delta_{\rm P_2}\simeq0.92\Gamma-1.5$. Repeated period doubling leads
towards the strongly chaotic region, whose large-scale boundary is
approximately described by
$\Delta_{\rm chaos}\simeq0.75\Gamma-3.5$. Within the chaotic region,
stable periodic windows form the characteristic shrimp-like structures
and exhibit an approximate period-adding sequence. The centres of the
most prominent structures follow
$\Delta_{\rm shrimp}\simeq0.416\Gamma-0.52$.

The nearly linear form of these boundaries at large $\Gamma$ shows that
the dynamics is governed mainly by
$ r\equiv \Delta/2\Gamma$.
In the limit $\epsilon =1/\Gamma\rightarrow0$, we reduce the continuous
three-mode equations to a one-dimensional slow--fast return map. This
map reproduces the principal period-doubling sequence and its transition
to chaos, with the $P_1\rightarrow P_2$, $P_2\rightarrow P_4$, and
$P_4\rightarrow P_8$ transitions approaching
$r\simeq0.466$, $0.400$, and $0.385$, respectively. Their accumulation
near $r\simeq0.38$ agrees closely with the large-$\Gamma$ limit
$r_{\rm chaos}\simeq0.375$ inferred from the full system. Deep minima of
the return-map Lyapunov exponent likewise coincide with the limiting
locations of the periodic windows in the full equations. The complicated
two-parameter bifurcation diagram can therefore be understood, to leading
order at large $\Gamma$, in terms of a much simpler one-dimensional
mapping. We also find Pomeau--Manneville type-I intermittency near the boundaries
of periodic windows. 

Allowing unequal daughter damping rates does not
remove the overall bifurcation structure, although both the Hopf and
period-doubling boundaries become progressively shallower as
$s=\Gamma_3/\Gamma_2$ decreases, reducing the extent of the chaotic
region in the $(\Gamma,\Delta)$ plane.

These results have a direct application to the outbursts observed in
cool DAV white dwarfs. In the three-mode interpretation of
\citet{Luan18}, an overstable parent $g$ mode slowly accumulates energy
and transfers it rapidly to a pair of strongly damped daughter modes,
whose dissipation produces an outburst. Estimates from DAV growth and
damping rates suggest physically relevant values
$\Gamma\sim10$--$30$, while strongly resonant daughter pairs can have
$\Delta\sim{\cal O}(1)$--${\cal O}(10)$. The two examples considered by LG18 are well inside the strongly nonlinear, chaotic part of our parameter space.
Our results therefore refine the usual criterion
$|\delta\omega|<|\gamma^E_{\rm d}|$: this condition does not imply a
single regular limit cycle, but encompasses simple periodic cycles,
period-doubled states, chaotic solutions, and narrow higher-period
windows. The observed diversity and aperiodicity of DAV outburst
recurrence may consequently arise naturally from modest differences in
mode detuning and damping rates among resonant triplets.

The isolated three-mode system studied here is the simplest building
block of a more general nonlinear mode network. Real pulsating stars may
contain several simultaneously active daughter pairs and interconnected
resonant triads, as well as additional effects from rotation,
non-adiabaticity, and higher-order nonlinearities. Extending the present
Lyapunov, continuation, and asymptotic analysis to such mode networks
will be important for connecting nonlinear bifurcation structure to observed stellar oscillations.


\appendix

\section{Floquet analysis and the period-doubling boundaries}
\label{sec:floquet}

To determine where a periodic solution becomes unstable and gives rise to a period-doubled solution, we use Floquet analysis. The basic idea is to follow a very small perturbation to a periodic orbit over one complete oscillation period and ask whether that perturbation grows or decays.

For example, to locate the transition from the fundamental periodic solution ($P_1$) to the period-doubled solution ($P_2$), we first calculate the $P_1$ periodic orbit $\boldsymbol{x}_{P1}(t)$ and its period $T$ using a shooting method. We then perturb this orbit slightly,
\begin{equation}
    \boldsymbol{x}(t)=\boldsymbol{x}_{P1}(t)+\delta\boldsymbol{x}(t),
\end{equation}
where $\delta\boldsymbol{x}$ is assumed to be small. To linear order, the perturbation evolves according to
\begin{equation}
    \frac{{\rm d}}{{\rm d}t}\delta\boldsymbol{x}
    =
    \mathbf{J}\!\left[\boldsymbol{x}_{P1}(t)\right]\delta\boldsymbol{x},
\end{equation}
where $\mathbf{J}$ is the Jacobian of the nonlinear amplitude equations evaluated along the periodic orbit.

We integrate these perturbation equations over one complete period $T$. The result can be written as
\begin{equation}
    \delta\boldsymbol{x}(T)
    =
    \mathbf{M}\,\delta\boldsymbol{x}(0),
\end{equation}
where $\mathbf{M}$ is the monodromy matrix. Its eigenvalues, called Floquet multipliers, describe how small perturbations change after one cycle. A multiplier with $|\mu|<1$ corresponds to a perturbation that decays from cycle to cycle, whereas $|\mu|>1$ indicates an instability.

One Floquet multiplier is always equal to unity because shifting a periodic solution slightly forward or backward in time produces the same orbit. The remaining, non-trivial multipliers determine the stability of the periodic solution.

A period-doubling, or flip, bifurcation occurs when a real non-trivial Floquet multiplier crosses the unit circle through
\begin{equation}
    \mu=-1.
\end{equation}
The minus sign means that the perturbation reverses sign after one period of the original orbit and therefore returns to the same sign only after two periods. This produces a new periodic solution whose period is twice that of the original one.

For the $P_1\rightarrow P_2$ transition, the Floquet multipliers are therefore calculated over the period $T$ of the $P_1$ orbit, and the boundary is located from $\mu_{P_1}=-1$. The same procedure applies to all subsequent period-doubling transitions. For the $P_2\rightarrow P_4$ boundary, we first calculate the stable $P_2$ orbit, whose period is approximately $2T$, and construct its monodromy matrix by integrating the variational equations over this full $P_2$ period. The transition to $P_4$ occurs when one of its non-trivial Floquet multipliers satisfies $ \mu_{P_2}=-1$.

As an independent calculation of the period-doubling boundaries, we also use \textsc{MatCont} to perform two-parameter numerical continuation in the $(\Delta,\Gamma)$ plane. \textsc{MatCont} identifies a period-doubling bifurcation using the same Floquet condition described above: a non-trivial multiplier of the periodic orbit reaches $\mu=-1$. It formulates the periodic orbit as a boundary-value problem using orthogonal collocation and directly continues the corresponding period-doubling curve as both $\Delta$ and $\Gamma$ are varied. The resulting curves therefore provide an independent numerical determination of the period-doubling boundaries.

\section{Type-I intermittency near periodic boundary points}
\label{sec:boundary}

For $\Delta=5$, we examined the transition near $\Gamma_{\rm c}\simeq21.638815$ (marked as a green circle in Fig. 2), where the regular solution is a $P_2$
limit cycle. On the intermittent side of the boundary, $\Gamma<\Gamma_{\rm c}$, the time series consists of long nearly
periodic laminar phases interrupted by irregular bursts (Fig. 1, right panel). Defining
\begin{equation}
    \epsilon=\frac{\Gamma_{\rm c}-\Gamma}{\Gamma_{\rm c}},
\end{equation}
we find that the maximum laminar duration obeys
\begin{equation}
    N_{\max}\propto \epsilon^{-0.484},
\end{equation}
in close agreement with the $\epsilon^{-1/2}$ scaling expected for
type-I intermittency \citep{Bussac1982,Meunier1982}. 

The dynamical origin of the intermittency is also visible in the
Poincar\'e return map. Since the underlying periodic orbit is $P_2$, we
consider the second-return map, $z_{n+2}$ versus $z_n$, which develops
a narrow near-tangent channel close to the bifurcation. Trajectories
spend many iterations drifting slowly through this channel before being
ejected into a burst, as expected for type-I intermittency. The largest
Lyapunov exponent changes from positive on the intermittent side to
approximately zero on the periodic side (Fig 2). In addition, we characterize the intermittency spectrally by constructing
a binary time series $I(t)$ that takes the value $I=0$ during the
near-periodic laminar phases and $I=1$ during the irregular bursts.
The low-frequency power spectrum of $I(t)$ is broadly consistent with
the logarithmically corrected type-I form \citep{Huang2023}:
\begin{equation}
    S_I(\nu)\propto
    \frac{1}{\nu |\log_{10}\nu|^2},
\end{equation}
over an intermediate frequency range of $\nu$, providing an additional diagnostic of the intermittent dynamics.

For $\Delta=2$, we find a second intermittent transition at
$\Gamma_{\rm c}\simeq16.827263$ (marked as an orange circle in Fig. 2). In this case the periodic attractor on
the regular side is a $P_3$ orbit. As the boundary is approached from the intermittent side, the time series again
shows increasingly long nearly periodic episodes separated by chaotic bursts. The measured laminar-length scaling is $N_{\max}\propto\epsilon^{-0.519}$,
while an alternative measure based on the longest complete laminar
episode gives
\begin{equation}
    L_{\max}\propto\epsilon^{-0.54}.
\end{equation}
Both exponents are close to the characteristic $-1/2$ scaling of type-I
intermittency. The same conclusion holds for the mean laminar length $ \langle L\rangle$ as well.

It is also useful to ask whether the intermittent behaviour could
instead arise through the Pomeau--Manneville type-II mechanism.
For a periodic orbit, the corresponding local bifurcation is a
Neimark--Sacker (torus) bifurcation, in which a complex-conjugate
pair of non-trivial Floquet multipliers crosses the unit circle.

For a periodic orbit of period $T$ in the symmetric three-mode
system ($s=1$), one Floquet multiplier is the trivial multiplier
$\mu_0=1$ associated with time translation, while the remaining
two multipliers are denoted by $\mu_1$ and $\mu_2$. The product $\mu_1\mu_2$  is
related to the phase-space volume change over one period. In
particular, for the variational system
$\dot{\delta\boldsymbol{x}}
=
\mathbf{J}(t)\,\delta\boldsymbol{x}$,
the determinant of the monodromy matrix satisfies
\[
\det\mathbf{M}
=
\exp\left[
\int_0^T {\rm tr}\,\mathbf{J}(t)\,dt
\right]=\mu_0\mu_1\mu_2.
\]
Since the trivial multiplier
$\mu_0=1$, the remaining two satisfy
\begin{equation}
    \mu_1\mu_2
    =
    \exp\left[2(1-\Gamma)T\right].
\end{equation}


For $\Gamma>1$ this product is smaller than unity. If
$\mu_1$ and $\mu_2$ form the complex-conjugate pair required for
a Neimark--Sacker bifurcation, then
$\mu_2=\mu_1^*$ and hence
\[
    |\mu_1|^2=|\mu_2|^2
    =\mu_1\mu_2
    =\exp\left[2(1-\Gamma)T\right] < 1.
\]
The pair therefore cannot cross the unit circle, excluding a
Neimark--Sacker bifurcation and hence the associated
Pomeau--Manneville type-II intermittency mechanism in the
symmetric system with $\Gamma>1$.


\section{Unequal daughters}
\label{sec:unequal}

\begin{figure}
\centering
\includegraphics[width=0.98\columnwidth]{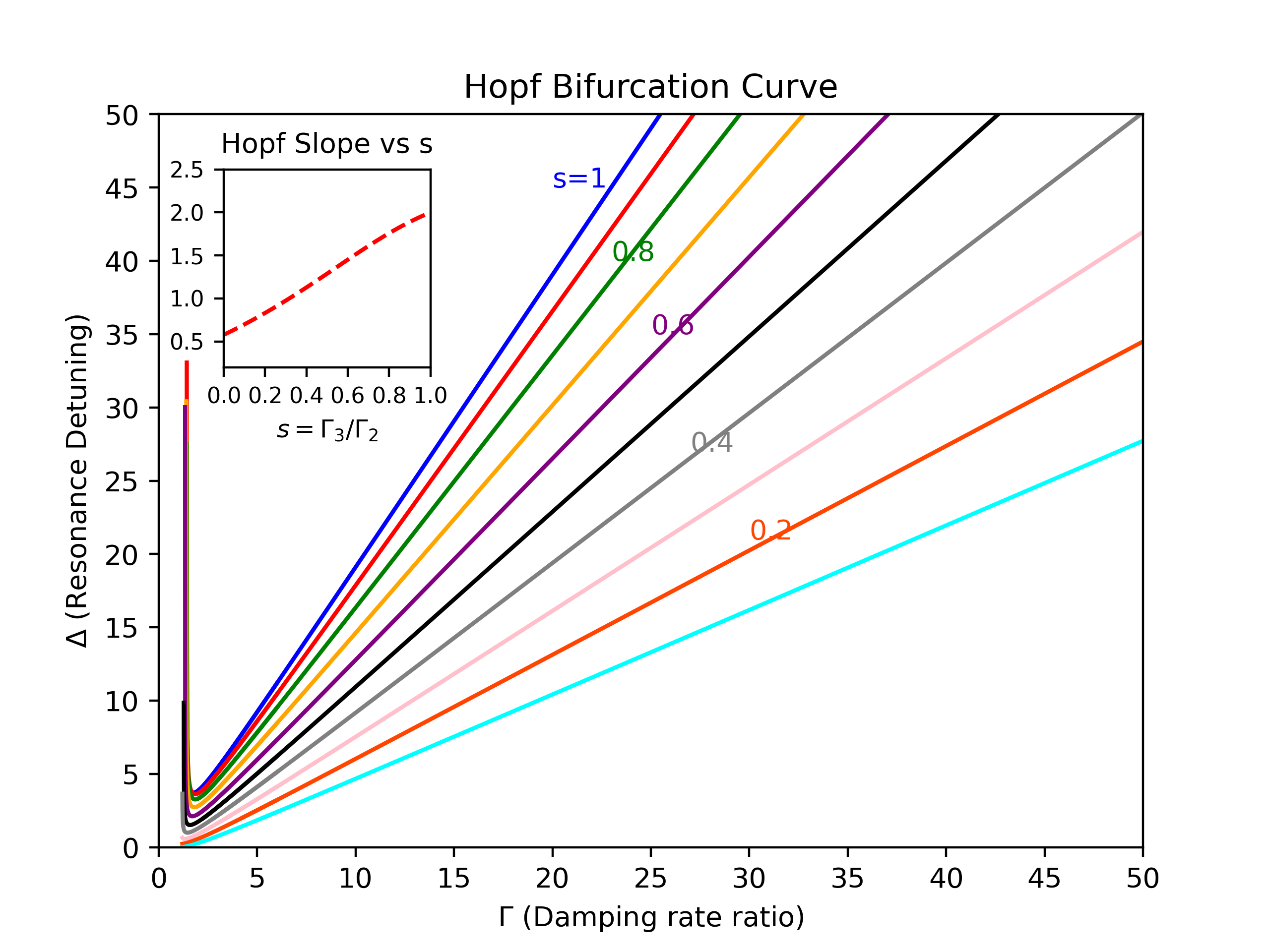}
\vspace{2cm}
\includegraphics[width=0.98\columnwidth]{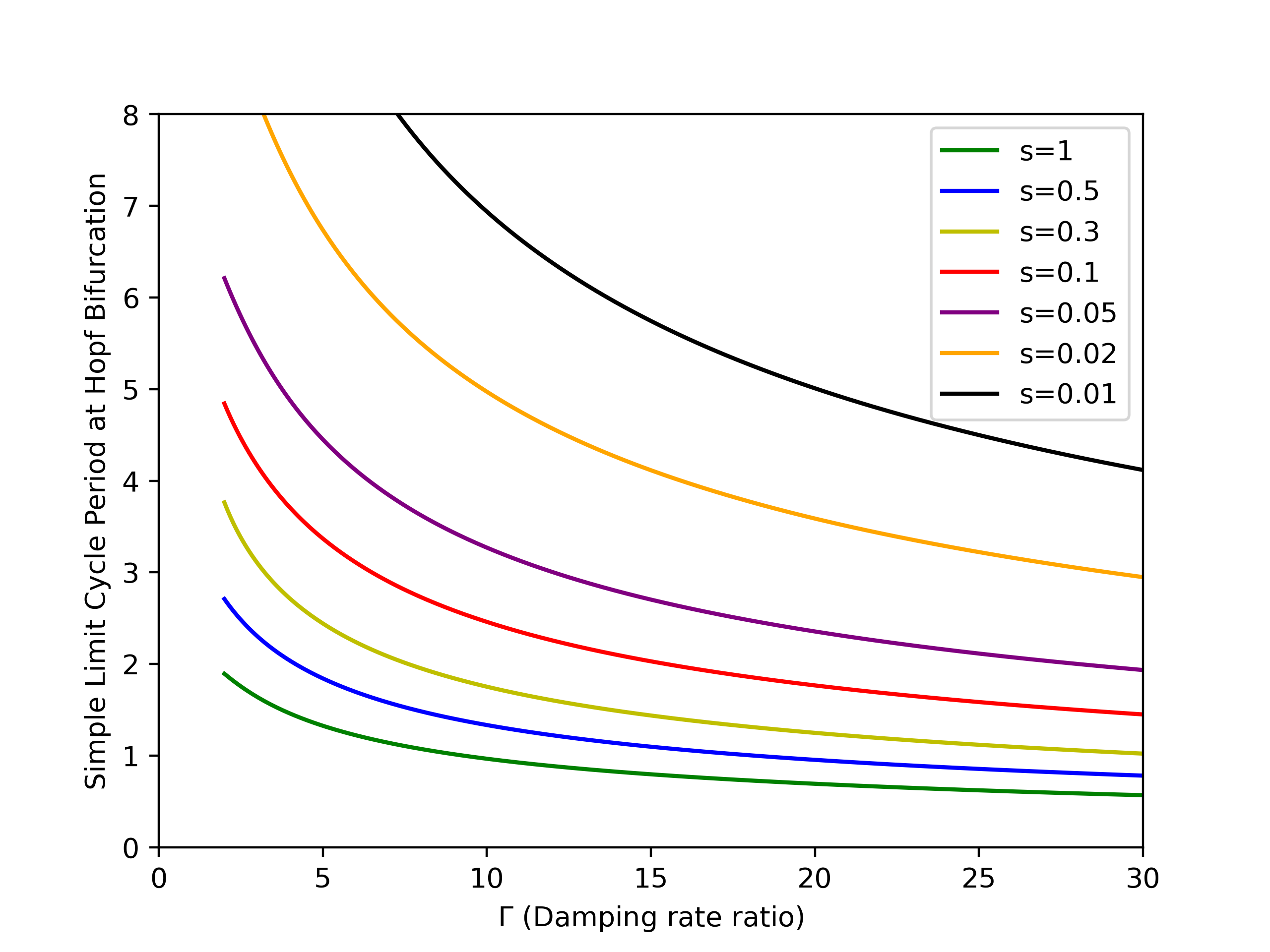}
\caption{Hopf-bifurcation properties of the three-mode coupling system for different
daughter-mode damping-rate ratios $s=\Gamma_3/\Gamma_2$.
The upper panel shows the equilibrium-to-$P_1$ limit-cycle boundary in the
$(\Gamma,\Delta)$ plane. As the daughter damping rates become more unequal
(i.e. as $s$ decreases from unity), the Hopf boundary becomes progressively
shallower. The inset shows the corresponding large-$\Gamma$ slope of the
Hopf boundary as a function of $s$. The lower panel shows the period of the
$P_1$ limit cycle at the Hopf bifurcation as a function of $\Gamma$ for
selected values of $s$. At fixed $\Gamma$, the oscillation period increases
as the damping-rate asymmetry becomes stronger.}
\label{fig:hopf_s}
\end{figure}

For a general $s=\Gamma_3/\Gamma_2$, we can obtain the $\Delta_{P_1}$ boundary (Hopf bifurcation) by solving for $\Delta'$ from Eq. \eqref{stability},
\begin{equation}
   a\Delta'^4+b\Delta'^2+c=0
\end{equation}
where $a = 3\left\{ (\Gamma(s+1) - 1)\left[ (\Gamma(1-s))^2 + 2(\Gamma(1+s))^2 + 1 \right] - 6s\Gamma^2 \right\}$,
$b = \left\{ (\Gamma(s+1) - 1)\left[ (\Gamma(1-s))^2 + (\Gamma(1+s))^2 + 2 \right] - 12s\Gamma^2 \right\}$, $c = -\left[ (\Gamma(s+1) - 1)^3 + 2s\Gamma^2 \right]$, and $\Delta' =\Delta/(1+\Gamma_2+\Gamma_3)$.

Expand the root for $\Delta$ as series as $\Gamma \rightarrow +\infty $,
We can obtain the asymptotic slope when $\Gamma$ is large, 
\begin{equation}
r' =\frac{\Delta}{\Gamma}=\frac{(1+s)}{\sqrt{3}(1-s)}\sqrt{  (2\sqrt{1-s^2+s^4}-(1+s^2)) } 
\end{equation}
which agrees very well with the numerical results shown in Fig.~\ref{fig:hopf_s}.
This slope approaches to $2+(s-1) + O((s-1)^2)$ as $s \rightarrow 1$, and
for small s it approaches to $1/\sqrt{3}+2s/\sqrt{3} + O(s^2)$.
The $\Delta_{P_1}$ vs. $\Gamma$ boundaries for different daughter mode ratios $s$ are shown in the upper panel of Fig.~\ref{fig:hopf_s}. The fitted $\Delta_{P_1}$ slope is overplotted in the inset. As \(s\) decreases from 1.0, the slope becomes progressively shallower. The corresponding period of $P_1$ limit cycle is plotted in the lower panel. 

\begin{figure}
\centering
\includegraphics[width=0.98\columnwidth]{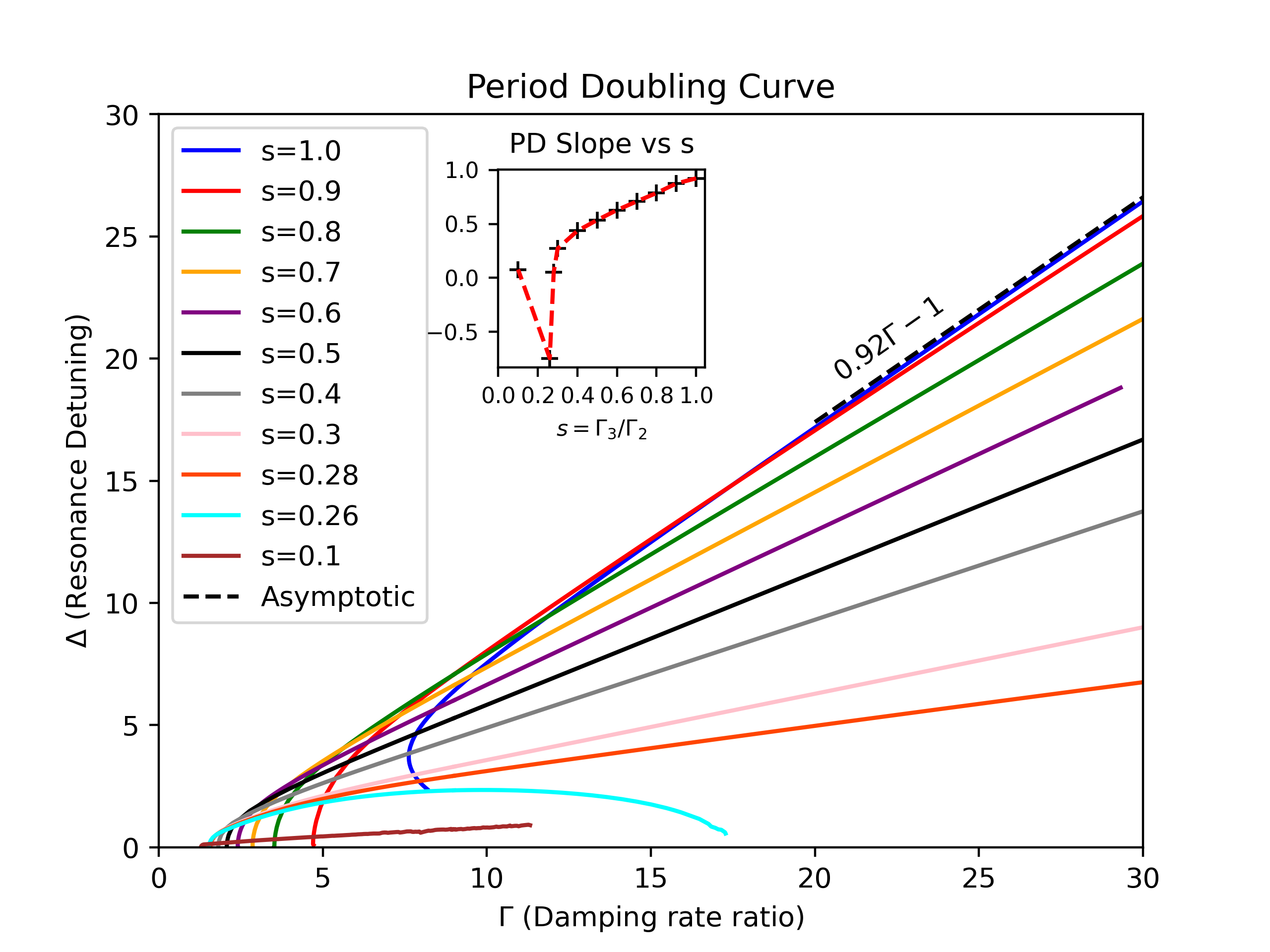}
\caption{$P_1\rightarrow P_2$ period-doubling boundaries in the
$(\Gamma,\Delta)$ plane for different daughter-mode damping-rate ratios
$s=\Gamma_3/\Gamma_2$. As $s$ decreases from unity, the boundary becomes
progressively shallower, indicating that the onset of period doubling shifts
to smaller detuning at fixed $\Gamma$. The dashed black line shows the
large-$\Gamma$ asymptotic relation for the symmetric case, while the inset
shows the fitted slope of the period-doubling boundary as a function of $s$.}
\label{fig:PD_s}
\end{figure}

Fig.~\ref{fig:PD_s} shows the period doubling $\Delta_{P_2}$ boundaries for different $s$. These $P_1 \rightarrow P_2$ boundaries are calculated by numerical continuation on the $(\Gamma,\Delta)$ plane using \textsc{MatCont}.
We also find a clear trend in which decreasing $s$ produces a shallower $\Delta_{P_2}$ boundary. This suggests that the chaotic region occupies a progressively smaller portion of parameter space on the $(\Gamma,\Delta)$ plane as the damping rates of the two daughter modes become more unequal.

\begin{figure}
\centering
\includegraphics[width=0.98\columnwidth]{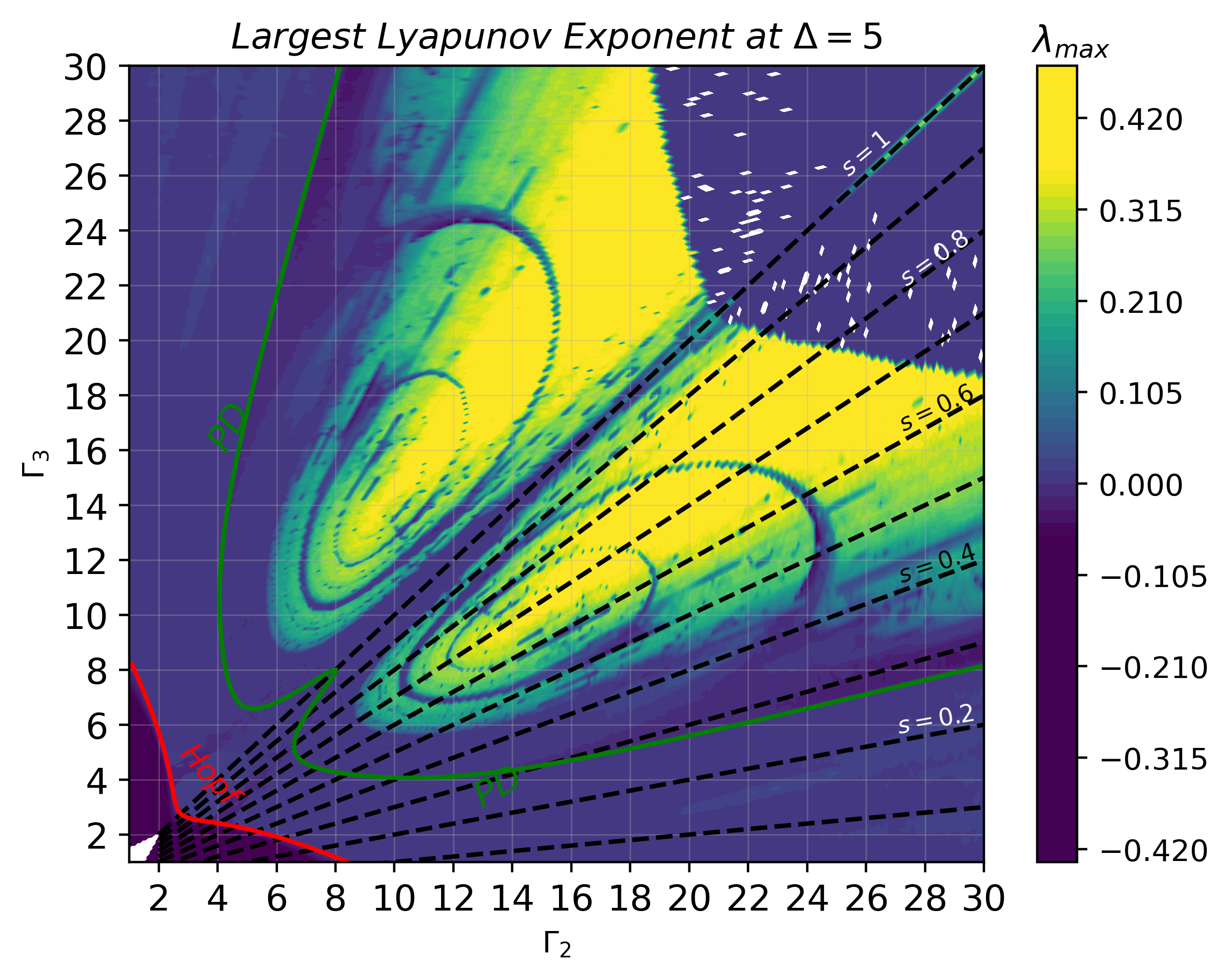}
\caption{Largest Lyapunov exponent $\lambda_{\rm max}$ in the
$(\Gamma_2,\Gamma_3)$ plane for fixed detuning $\Delta=5$.
The red curve marks the Hopf bifurcation at which the stationary
equilibrium loses stability and a $P_1$ limit cycle emerges, while the
green curve marks the first period-doubling boundary
$P_1\rightarrow P_2$. Dashed black lines indicate constant daughter-mode
damping-rate ratios $s=\Gamma_3/\Gamma_2$. The map reveals the broad
distribution of periodic and chaotic regimes as the two daughter damping
rates are varied independently.}
\label{fig:lyap_d5}
\end{figure}

Fig.~\ref{fig:lyap_d5} shows the largest Lyapunov exponent in the
$(\Gamma_2,\Gamma_3)$ plane for fixed detuning
$\Delta=5$. The colour scale distinguishes the main dynamical regimes:
negative values of $\lambda_{\max}$ correspond to stable equilibria,
$\lambda_{\max}\simeq 0$ to periodic attractors, and positive values to
chaotic solutions, with the light-green and yellow regions marking the
strongly chaotic part. The red curve denotes the
Hopf bifurcation boundary at which the equilibrium loses stability and a
$P_1$ limit cycle is created. The green curve labelled PD marks the first
period-doubling boundary, where the $P_1$ limit cycle becomes unstable and
a $P_2$ solution appears. Beyond this boundary, further bifurcations give
rise to alternating periodic windows and extended chaotic regions,
producing the intricate structure visible in the Lyapunov map. The black
dashed lines indicate constant daughter-mode damping-rate ratios
$ s \equiv \Gamma_3/\Gamma_2$,
with representative values of $s$ labelled in the figure, allowing the
dynamics to be followed along cuts of fixed relative damping between the
two daughter modes.

\section{Phase-rotational form of amplitude equations}
\label{sec:rotational}

For the numerical calculation of the Lyapunov exponents, we transform
the amplitude--phase system to Cartesian-like variables \citep{Rech2010,Hughes1990}
\[
X=a_1\sin\phi,\qquad
Y=a_1\cos\phi,\qquad
Z=a_2,\qquad
\beta=a_3 .
\]
The resulting equations are
\begin{align}
\dot{X} &= X+Z\beta+\Delta Y-BY^2,\\
\dot{Y} &= Y-\Delta X+BXY,\\
\dot{Z} &= -\Gamma_2 Z-X\beta,\\
\dot{\beta} &= -\Gamma_3\beta-XZ,
\label{phaseAE}
\end{align}
where
\[
B=\frac{Z^2+\beta^2}{Z\beta}.
\]
For equal daughter damping rates, $\Gamma_2=\Gamma_3=\Gamma$, and
symmetric initial conditions, $Z=\beta$ defines an invariant
submanifold, for which $B=2$. The equations then reduce to 
\begin{align}
\dot{X} &= X+Z^2+\Delta Y-2Y^2,\\
\dot{Y} &= Y-\Delta X+2XY,\\
\dot{Z} &= -(\Gamma+X)Z.
\end{align}

\section{1D return map}
\label{sec:return}
\begin{figure}
   \centering    
 \includegraphics[width=8cm]{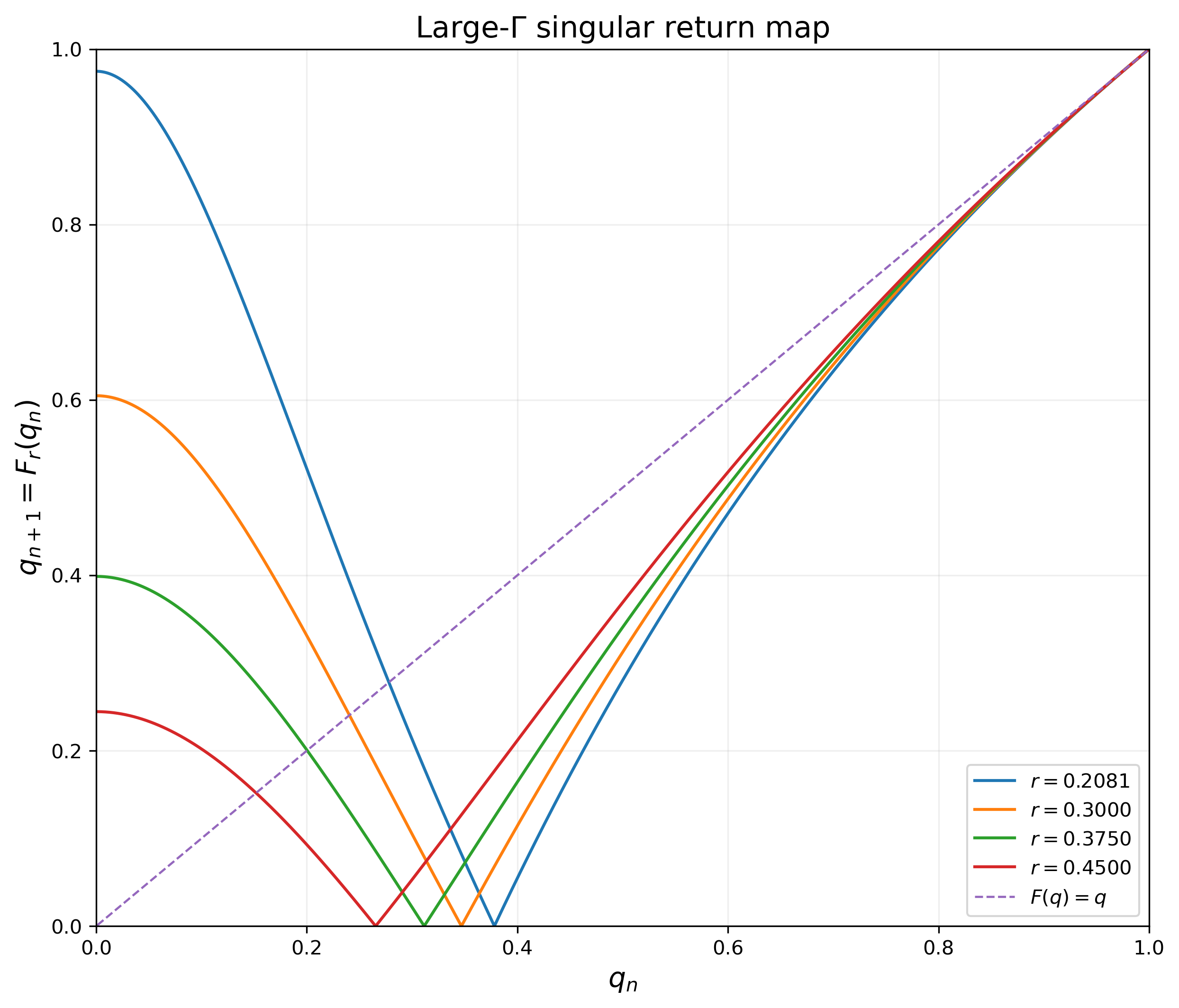}
   \caption{1D return map $q_{n+1}=F_r(q_n)$ for different $r$. The dotted line indicates the $F_r(q)=q$ level.}
\label{1dmap}%
 \end{figure}

We plot the 1D return map  $q_{n+1}=F_r(q_n)$ discussed in Section 4 with different $r$ values in Fig.~\ref{1dmap}.

\section*{Acknowledgments}

We are grateful to Phil Arras, Nevin Weinberg and Adrian Barker for insightful discussions. This work was supported by STFC grants ST/W000873/1 and UKRI1179. ZG used ChatGPT (OpenAI) to assist with language editing and the derivation of the 1D return map. All AI-assisted content was reviewed and verified by the authors, who take full responsibility for the final manuscript and the scientific results presented therein.

\section*{Data availability}
The numerical scripts and data products supporting the figures 
will be made available on Zenodo with the published version of this work.


\bibliographystyle{mnras}
\bibliography{coupleMN} 





\bsp
\label{lastpage}

\end{document}